\pdfoutput=1
\documentclass[pra,nofootinbib,twocolumn,amsmath,amssymb,superscriptaddress]{revtex4-2}
\usepackage{graphicx,amsmath,relsize,epstopdf,color,mathtools,bm,newtxtext,newtxmath,braket,rotating}
\usepackage[hyphenbreaks]{breakurl}
\usepackage[colorlinks=true,linkcolor=blue,citecolor=blue,urlcolor =blue]{hyperref}
\usepackage[normalem]{ulem}

\usepackage{booktabs}
\usepackage{multirow}

\usepackage[table,xcdraw]{xcolor}
\newcommand{\eq}[1]{\begin{equation}\begin{aligned}#1\end{aligned}\end{equation}}
\newcommand{\expct}[1]{\left\langle#1\right\rangle}
\newcommand{\Cov}{\mathop{\mathrm{Cov}} \nolimits}
\newcommand{\Var}{\mathop{\mathrm{Var}} \nolimits}

\newcommand{\iu}{\text{i}}
\newcommand{\eu}{\text{e}}
\newcommand{\ha}{\hat{a}}
\newcommand{\had}{\hat{a}^\dagger\vphantom{a}}
\newcommand{\hb}{\hat{b}}
\newcommand{\hbd}{\hat{b}^\dagger\vphantom{a}}

\usepackage{soul,xcolor}

\usepackage{dsfont}

\begin{document}
	\setstcolor{red}

	\title{Multiparameter transmission estimation at the quantum Cram\'er-Rao limit on a cloud quantum computer}
	
	\author{Aaron Z. Goldberg}
	\affiliation{National Research Council of Canada, 100 Sussex Drive, Ottawa, Ontario K1N 5A2, Canada}
	\affiliation{Department of Physics, University of Ottawa, Advanced Research Complex, 25 Templeton Street, Ottawa,Ontario K1N 6N5, Canada}
	
	\author{Khabat Heshami}
	\affiliation{National Research Council of Canada, 100 Sussex Drive, Ottawa, Ontario K1N 5A2, Canada}
	\affiliation{Department of Physics, University of Ottawa, Advanced Research Complex, 25 Templeton Street, Ottawa,Ontario K1N 6N5, Canada}
	\affiliation{Institute for Quantum Science and Technology, Department of Physics and Astronomy, University of Calgary, Alberta T2N 1N4, Canada}

	\begin{abstract}
		Estimating transmission or loss is at the heart of spectroscopy. To achieve the ultimate quantum resolution limit, one must use probe states with definite photon number and detectors capable of distinguishing the number of photons impinging thereon. In practice, one can outperform classical limits using two-mode squeezed light, which can be used to herald definite-photon-number probes, but the heralding is not guaranteed to produce the desired probes when there is loss in the heralding arm or its detector is imperfect. We show that this paradigm can be used to simultaneously measure distinct loss parameters in both modes of the squeezed light, with attainable quantum advantages. We demonstrate this protocol on Xanadu's X8 chip, accessed via the cloud, building photon-number probability distributions from $10^6$ shots and performing maximum likelihood estimation (MLE) on these distributions $10^3$ independent times. Because pump light may be lost before the squeezing occurs, we also simultaneously estimate the actual input power, using the theory of nuisance parameters. MLE converges to estimate the transmission amplitudes in X8's eight modes to be 0.39202(6), 0.30706(8), 0.36937(6), 0.28730(9), 0.38206(6), 0.30441(8), 0.37229(6), and 0.28621(8) and the squeezing parameters, which are proxies for effective input coherent-state amplitudes, their losses, and their nonlinear interaction times, to be 1.3000(2), 1.3238(3), 1.2666(2), and 1.3425(3); all of these uncertainties are within a factor of two of the quantum Cram\'er-Rao bound. This study provides crucial insight into the intersection of quantum multiparameter estimation theory, MLE convergence, and the characterization and performance of real quantum devices.
	\end{abstract}
	
	\maketitle
\tableofcontents

\section{Introduction}

\begin{figure*}
    \centering
    \includegraphics[width=\textwidth]{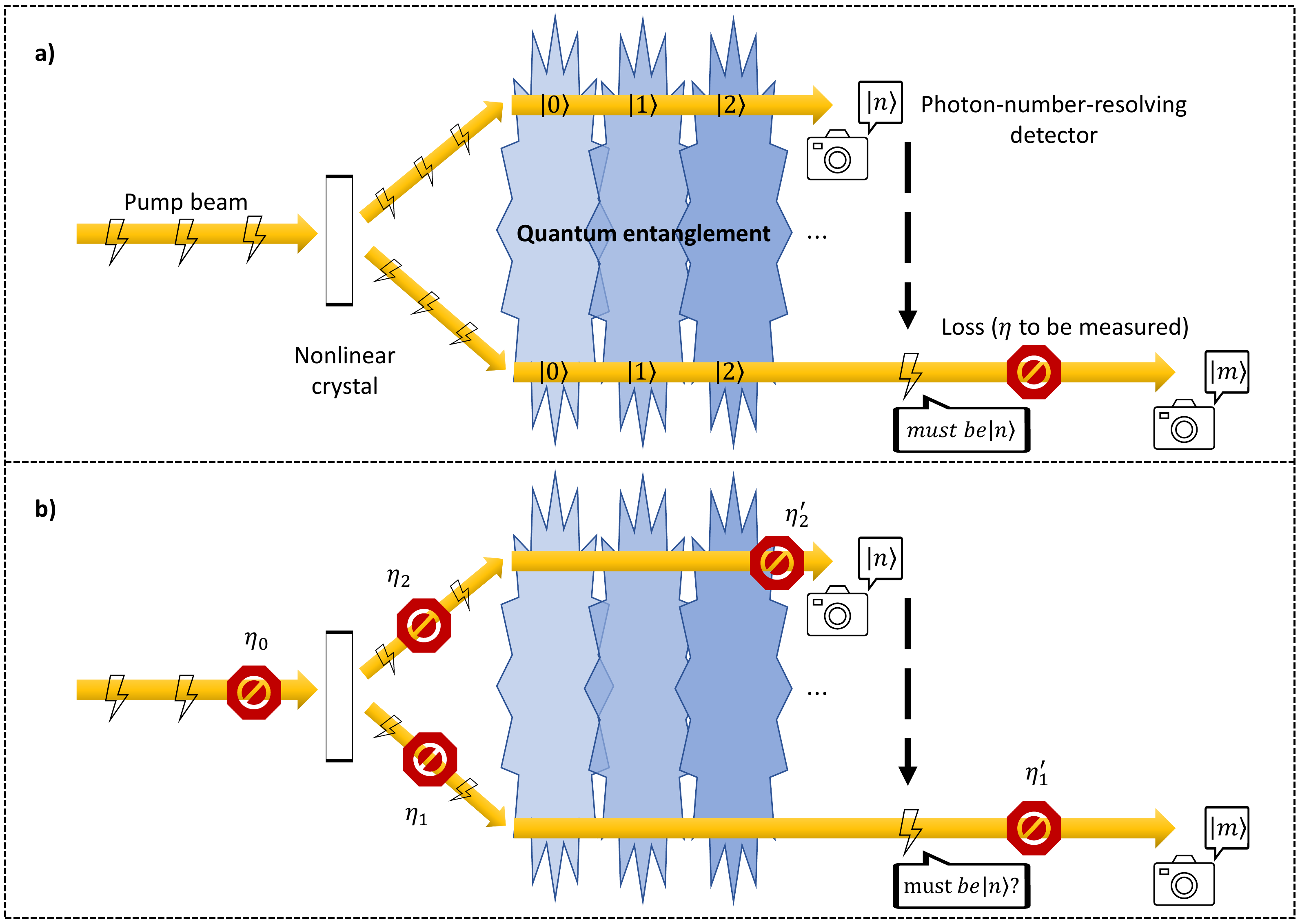}
    \caption{\textbf{a)} Optimal and \textbf{b)} realistic schemes for estimating any parameter related to transmission or loss. \textbf{a)} In the optimal scheme, a two-mode squeezed vacuum (TMSV) state with photon-number entanglement between its two arms is generated by impinging a pump beam on a nonlinear crystal. Measuring the photon number of one arm at a photon-number-resolving detector (PNRD) collapses the other arm into a definite, known, photon-number state, which can then be used to optimally sense the transmission amplitude $\eta$ by counting the number of photons that are transmitted. \textbf{b)} In the realistic scheme, there can be loss in the pump beam before it generates entanglement or the generation can be imperfect ($\eta_0$), there can be loss in each arm of the TMSV ($\eta_1$ and $\eta_2$), there can be loss if the light is coupled into or out from various fibers or if there are other mode mismatches (also contained in $\eta_1$ and $\eta_2$), and the detectors can be inefficient, modeled by the loss of photons prior to the detection event  ($\eta_1^\prime$ and $\eta_2^\prime$). Then, a measurement of $n$ photons at one PNRD no longer guarantees the other mode to have $n$ photons. The relevant estimation procedure in any realistic scenario is thus to simultaneously estimate all of the transmission parameters along all three branches. For sufficiently large transmission amplitudes, TMSV and PNRDs provide a dramatic quantum advantage in this simultaneous estimation procedure.}
    \label{fig:schematic}
\end{figure*}
Quantum probes and measurement devices are capable of estimating parameters with more precision than their classical counterparts, given a fixed amount of resources \cite{Caves1981,Dowling1998,Giovannettietal2004,Mitchelletal2004,Berryetal2009,LIGO2011,Humphreysetal2013,Tayloretal2013,Tsangetal2016,Liuetal2020,LupuGladsteinetal2022}. One particular class of measurements investigates the amount of light transmitted through, absorbed by, or reflected from a material of interest: this has applications ranging from ellipsometry \cite{AzzamBashara1977,Fujiwara2007,Tuchin2016} to spectroscopy \cite{Cheongetal1990,Savageetal1996,Hollas2004,Coneetal2015} to ghost imaging \cite{ErkmenShapiro2010} to characterization of quantum devices \cite{Spagnoloetal2014,Wangetal2017,Arrazolaetal2021}. In these circumstances, specially designed quantum probes and detectors can decrease the parameter estimate's variance by the fraction of light lost \cite{JakemanRarity1986,Heidmannetal1987,Hayatetal1999,Abouraddyetal2001,Abouraddyetal2002entangledellipsometry,Toussaintetal2004,Grahametal2006,MonrasParis2007,Brambillaetal2008,Adessoetal2009,MonrasIlluminati2011,Alipouretal2014,Crowleyetal2014,Medaetal2017,Loseroetal2018,Nair2018,Rudnickietal2020,Ioannouetal2021,WangAgarwal2021}, due to the increased correlations in their photon-number distributions, which is particularly useful in the sensing of increasingly faint signals such as when a trace amount of a substance absorbs a small amount of light from a particular electromagnetic field mode. These quantum advantages have been demonstrated using either single photons \cite{YabushitaKobayashi2004,Bridaetal2010,Moreauetal2017,SabinesChesterkingetal2017,Samantarayetal2017,Whittakeretal2017,Yoonetal2020} or squeezed light \cite{Heidmannetal1987,Tapsteretal1991,SoutoRibeiroetal1997,DAuriaetal2006,Brambillaetal2008,Loseroetal2018,Shietal2020,Atkinsonetal2021,Woodworthetal2022} as probe states.

Many real-world measurement problems involve determining the values of multiple parameters. Multiparameter estimation theory is especially intricate in the quantum realm \cite{Paris2009,TothApellaniz2014,Szczykulskaetal2016,Braunetal2018,Albarellietal2020,DemkowiczDobrzanskietal2020,Polinoetal2020,SidhuKok2020,Goldbergetal2021singular,Liuetal2022}, with perils such as parameter incompatibility \cite{Matsumoto2002,Zhu2015, Heinosaarietal2016, Ragyetal2016} and boons such as precision enhancements arising from the simultaneous versus sequential estimation of the parameters \cite{Humphreysetal2013,BaumgratzDatta2016,Goldbergetal2020multiphase,Houetal2020,Goldbergetal2021intrinsic,GoreckiDemkowiczDobrzanski2022}. Our goal is to simultaneously determine the amount of light transmitted through two separate modes using a single quantum resource state: two-mode squeezed vacuum (TMSV). We find that we must also estimate the input power of the pump creating the TMSV, so this is a natural three-parameter estimation problem that is markedly different from three copies of a single-parameter estimation problem, even though the parameters are in principle all compatible with one another.

Squeezed light has often been used in single-parameter estimation, especially, in the context of transmission measurements, as a heralded source of single photons \cite{Bridaetal2011,Krapicketal2013} that can then be used as probe states. In the bright regime, it can be used as a heralded source of Fock states with definite photon number, which are optimal for sensing transmission or loss \cite{Adessoetal2009}. However, this heralding comes with the caveat of requiring low levels of loss in the heralding mode \cite{Moreauetal2017,Woodworthetal2020,Dowranetal2021}, which also requires perfectly efficient heralding detectors. Without these requirements, one cannot guarantee that the probe states are the Fock states that one desires.
We relax these requirements to see what information can be gleaned about the loss in both modes without any previous knowledge of either mode's transmission probabilities and allowing for loss in the TMSV generation process (see Fig. \ref{fig:schematic}). Our theoretical demonstrations certify, using the quantum Fisher information matrix (QFIM), that it is possible to outperform classical sensing paradigms with comparable probe energies, with the advantage increasing with increasingly large transmission probabilities.

We demonstrate our protocol on Xanadu's X8 integrated photonic quantum computer \cite{Arrazolaetal2021}, which provides both the TMSV probe states and the photon-number-resolving detectors (PNRDs) crucial to this quantum sensing advantage. Xanadu's cloud-accessible quantum computers have recently been demonstrated to achieve landmark \textit{computational} advantages using squeezing and PNRDs \cite{Madsenetal2022} and we aim to demonstrate \textit{metrological} advantages with these fruitful resources. This, further, serves as an opportunity to use quantum sensing techniques for characterizing \textit{quantum} devices; we do not simulate loss with any \textit{a priori} technique but, rather, measure the overall fraction of light transmitted through the entire device.  Other technologies developed for quantum computation \cite{Chengetal2022arxiv} will similarly be useful for quantum metrology in the immediate future.

There are important statistical considerations that need be especially rigorous when trying to demonstrate a quantum advantage. One avenue we study is the effect of varying pump powers on the estimation of the loss parameters; when this effect is neglected, the statistical predictions differ dramatically in the low-transmission regime, as is well known from studies of so-called ``nuisance parameters'' \cite{Basu1977,Suzuki2020,Suzukietal2020}.
Our final contribution in this work is to provide a comparison of different approaches to maximum likelihood estimation (MLE) with our data. Depending on the details of the technique used, we can show a quantum advantage, outperform the optimal \textit{quantum} limit, saturate the Cram\'er-Rao bound, or \textit{underperform} relative to classical precision limits. Given our physical setup, we show that TMSV probes are \textit{insufficient} to beat the classical estimation limit when the loss probabilities and input energies are this high, and explain the discrepancies between different statistical techniques. Since we approximately saturate the Cram\'er-Rao precision bound for this multiparameter setup, we can promisingly conclude that the X8 device could certainly demonstrate a quantum multiparameter metrology advantage in the easier scenario of providing \textit{less intense} squeezed light. The lessons to be learnt include the parametric dependence of quantum advantages, the true cost of quantum advantages in terms of interrogation time required on fragile machines, and the caveats of using a statistical quantity whose properties are only guaranteed in the asymptotic limit.

\section{Theory}
\subsection{Quantum limits on single-parameter loss estimation}
Consider a single bosonic mode with annihilation operator $\ha$. After impinging on a beam splitter with transmission amplitude $\eta$ along with another ``vacuum'' mode annihilated by $\hat{v}$, this operator transforms as
\eq{
    \ha\to \eta\ha+\sqrt{1-\eta^2}\hat{v}.
    \label{eq:input output amplitude}
} If the vacuum mode is initially unpopulated, this leads to the same reduction in intensity for all input states
\eq{
    \expct{\had \ha}\to \eta^2\expct{\had\ha}.
} As such, the input-output relation in Eq. \eqref{eq:input output amplitude} can be used to model a plethora of physical situations that lead to intensity reductions, ranging from loss through a noisy channel to incomplete reflection, depending on what one considers as the output mode. Notably, sequential applications of Eq. \eqref{eq:input output amplitude} lead to another input-output relation of the same form, so we can always model the \textit{total} loss or transmittance of a physical scenario by considering the parameter $\eta$ alone, encompassing everything from loss in a network to imperfect detectors.

Given this transformation, how can one best estimate $\eta$? Quantum estimation theory dictates that the variance in any estimate of $\eta$ will be lower-bounded by the inverse of the quantum Fisher information (QFI) $\mathsf{F}_{\hat{\rho}}(\eta)$, as encoded in the quantum Cram\'er-Rao bound (qCRB):
\eq{
    \Var_{\hat{\rho}}(\eta)\geq \frac{1}{\mathsf{F}_{\hat{\rho}}(\eta)}.
} We have here explicitly included the dependence on the probe state $\hat{\rho}$, for the bound can be improved or worsened by considering different probes, and made no reference to the detection procedure because the QFI automatically optimizes over all such protocols.
For a pure probe $\hat{\rho}=\ket{\psi}\bra{\psi}$, the QFI takes the form
\eq{
    \mathsf{F}_{\psi}(\eta)=4M \left(\braket{\partial_\eta \psi|\partial_\eta \psi}-\left|\braket{\partial_\eta \psi| \psi}\right|^2\right),
} where $M$ is the number of times the measurement is repeated. This provides a unique figure of merit for choosing an optimal probe state, so we can consider $M=1$ when optimizing such a state.

Classical light with an average number of photons $E$ (i.e., average energy $E$ in units of energy per photon) takes the form of a coherent state with $|\alpha|^2=E$:
\eq{
    |\alpha\rangle\propto \sum_{n=0}^\infty \frac{(\alpha \had)^n}{n!}\ket{\mathrm{vac}}.
} This state undergoes the simple transformation $|\alpha\rangle\to|\eta\alpha\rangle$ and thus has the QFI
\eq{
    \mathsf{F}_\alpha(\eta)=4|\alpha|^2=4E.
} This can be improved upon by using Fock states with definite photon number $E=N$:
\eq{
    \ket{E=N}=\frac{\had^N}{\sqrt{N!}}\ket{\mathrm{vac}}.
} Such states lose their purity when photons are lost, transforming as
\eq{
    \ket{N}\bra{N}\to\sum_{n=0}^N\binom{N}{n}\eta^{2n} \left(1-\eta^2\right)^{N-2}\ket{n}\bra{n},
} yet are proven to be optimal for sensing such loss, with QFI
\eq{
    \mathsf{F}_N(\eta)=\frac{4N}{1-\eta^2}.
} These present a particularly dramatic improvement in sensing with small amounts of loss $\eta\approx 1$.

How can these quantum limits be attained in practice? One method is to herald the presence of Fock states from easier-to-create TMSV states defined in two orthogonal bosonic modes annihilated by $\ha$ and $\hb$:
\eq{
    \ket{\mathrm{TMSV}}\propto 
    \sum_{n=0}^\infty \left(-\eu^{\iu\varphi}\tanh r \right)^n\ket{n}_a\otimes\ket{n}_b,
    \label{eq:TMSV}
} with average energy $\sinh^2 r$ in each mode. Then, if one uses a PNRD in mode $b$ and measures some number $N$ photons to be present, this heralds the presence of the Fock state $\ket{N}$ in mode $a$ that can then be used for estimation tasks. Moreover, the optimal measurement scheme to estimate $\eta$ is to use PNRDs in mode $a$, as the probability distribution $\left|\braket{N|\psi}\right|^2$ is the most sensitive possible for estimating $\eta$. The optimal scheme may thus consist of creating a TMSV, sending one mode directly to a PNRD, and having the second mode interact with the absorptive medium before also being sent to a PNRD (Fig. \ref{fig:schematic}\textbf{a)}). Such a setup relies on there being no loss in the mode sent directly to a PNRD, because otherwise the photon numbers in each mode are not always equal, which is the restriction we circumvent in this work. 

We note that different parametrizations can be used for estimating loss; some examples include the transmission probability $q=\eta^2$ and the parameter $\phi=\arccos \eta$ defined by the master equation $\partial_\phi \rho=\tan \phi \left(2\had \hat{\rho}\ha-\had\ha\hat{\rho}-\hat{\rho}\had\ha\right)$, which might arise more naturally in particular physical problems. The QFI for a new parametrization is simply a transformed version of the old, which we write in the suggestive form
\eq{
    \mathsf{F}_{\hat{\rho}}(\vartheta)=\frac{\partial \eta}{\partial \vartheta}\mathsf{F}_{\hat{\rho}}(\eta)\frac{\partial \eta}{\partial \vartheta},
    \label{eq: Jacobian transformation single parameter}
} so our results can easily be adapted to fit one's preferred quantification of loss.

\subsection{Quantum limits on multiparameter loss estimation}
Two modes experience loss in parallel versions of Eq. \eqref{eq:input output amplitude}:
\eq{
    \ha&\to \eta_1\ha+\sqrt{1-\eta_1^2}\hat{v}_1\\
    \hb&\to \eta_2\hb+\sqrt{1-\eta_2^2}\hat{v}_2.
    \label{eq:input output amplitude two mode}
} If these two modes correspond to two different polarization modes, for example, then this process defines a diattenuation; i.e., dichroism \cite{Goldberg2020}. 

In this multiparameter context, which can readily be expanded to any integer $k$ modes with transmission amplitudes $\eta_k$, one is interested in minimizing the covariances between the estimates of all $k$ parameters $\pmb{\eta}$. The covariance matrix is also governed by the qCRB, now bounded by the QFIM:
\eq{
    \Cov_{\hat{\rho}}(\pmb{\eta})\succeq \boldsymbol{\mathsf{F}}_{\hat{\rho}}(\pmb{\eta})^{-1},
    \label{eq:qCRB multiparameter}
} where the latter has components
\eq{
    \left[{\mathsf{F}}_{\psi}(\pmb{\eta})\right]_{ij}=4M\mathrm{Re}\left[\braket{\partial_{\eta_i} \psi|\partial_{\eta_j} \psi}-\braket{\partial_{\eta_i} \psi| \psi}\braket{ \psi|\partial_{\eta_j} \psi}\right]
} and matrix inequalities $\mathbf{A}\succeq\mathbf{B}$ imply that $\mathbf{A}-\mathbf{B}$ is positive semidefinite. One can change parametrizations by generalizing Eq. \eqref{eq: Jacobian transformation single parameter} to include Jacobians with components $J_{ij}=\partial \eta_i/\partial \vartheta_j$
\eq{
    \boldsymbol{\mathsf{F}}_{\hat{\rho}}(\pmb{\vartheta})=\mathbf{J}^\top\boldsymbol{\mathsf{F}}_{\hat{\rho}}(\pmb{\eta})\mathbf{J},
    \label{eq: Jacobian transformation multiparameter}
} where $^\top$ denotes the matrix transpose. An important difference from single-parameter estimation is that the qCRB in Eq. \eqref{eq:qCRB multiparameter} is not always guaranteed to be saturable.

Using classical light $\ket{\alpha}_a\otimes\ket{\beta}_b$ to probe these two parameters leads to a diagonal QFIM (again we take $M=1$)
\eq{
    \boldsymbol{\mathsf{F}}_{\alpha,\beta}(\eta_1,\eta_2)=4\begin{pmatrix} |\alpha|^2&0\\ 0&|\beta|^2
    \end{pmatrix}.
    \label{eq:QFIM coherent}
} To choose the optimal probe state, one must make a choice as to how to weigh the relative importance of minimizing $\Var(\eta_1)$ versus $\Var(\eta_2)$. This can often be chosen in a natural way \cite{Goldbergetal2021intrinsic} and, indeed, here one simply weighs the fraction of the total energy $|\alpha|^2+|\beta|^2$ put into each probe-state mode according to the relative weights given to the two aforementioned variances. With no other distinguishing information due to neither mode being prioritized, we will henceforth seek to minimize the total variance $\Var(\eta_1)+\Var(\eta_2)$, which, in this case, is optimized by putting half of the energy into the classical state in each mode.

Since the transformation of Eq. \eqref{eq:input output amplitude two mode} does not mix the two modes, it is clear that the optimal quantum sensing protocol is to repeat the optimal single-mode sensing protocols in parallel: use a probe state $\ket{m}_a\otimes\ket{n}_b$, impinge it on the sample, then measure the light using a PNRD for each mode. Again, the QFIM is diagonal, taking the form 
\eq{
    \boldsymbol{\mathsf{F}}_{m,n}(\eta_1,\eta_2)=4\begin{pmatrix} \frac{m}{1-\eta_1^2}&0\\ 0&\frac{n}{1-\eta_2^2}
    \end{pmatrix}.
    \label{eq:QFIM Fock}
} In this case, the optimal choice for how to apportion the energy between the two modes depends on the actual values of the parameters being estimated. As in the single-parameter case, a larger transmission amplitude leads to a greater quantum enhancement in sensitivity, so this can appropriately be balanced with the amount of energy in each mode. When no \textit{a priori} information is present, it behooves one to balance the probe states with $m=n$.

\subsection{Quantum limits on multiparameter loss estimation with TMSV}
It is difficult to produce Fock states with large numbers of photons. Most of the quantum-enhanced transmission estimation experiments have thus relied on TMSV as a source of nonclassical light, especially in the dim regime with $\sinh^2r \ll 1$. We next investigate the performance of TMSV for simultaneously estimating two loss parameters. This is especially pertinent when there is a physical relationship between the two modes being probed: for example, if one is interested in estimating the amount of attenuation in two orthogonal polarization modes to measure linear or circular dichroism, TMSV is a natural choice because it directly produces nonclassical (entangled) states within these two modes.

How does TMSV evolve under Eq. \eqref{eq:input output amplitude two mode}? It evolves into the mixed state \cite{Goldberg2022}
\eq{
    \hat{\rho}(\eta_1,\eta_2)=\sum_{m,n=0}^\infty \left(\frac{1-\eta_1^2}{\eta_1^2}\right)^m\left(\frac{1-\eta_2^2}{\eta_2^2}\right)^n\ket{\psi_{m,n}}\bra{\psi_{m,n}},
    \label{eq:rho after TMSV loss}
} where we have defined the (unnormalized) states
\eq{
    \ket{\psi_{m,n}}=\frac{1}{\cosh r}\sum_{N\geq m,n} f^N \sqrt{\binom{N}{m}\binom{N}{n}}\ket{N-m}_a\otimes\ket{N-n}_b
}
and the parameter-dependent function
\eq{
f(\eta_1,\eta_2;r,\varphi)=-\eu^{\iu\varphi}\eta_1\eta_2\tanh r.
}

\subsubsection{Two-parameter estimation}
\begin{figure}
    \centering
    \includegraphics[width=\columnwidth]{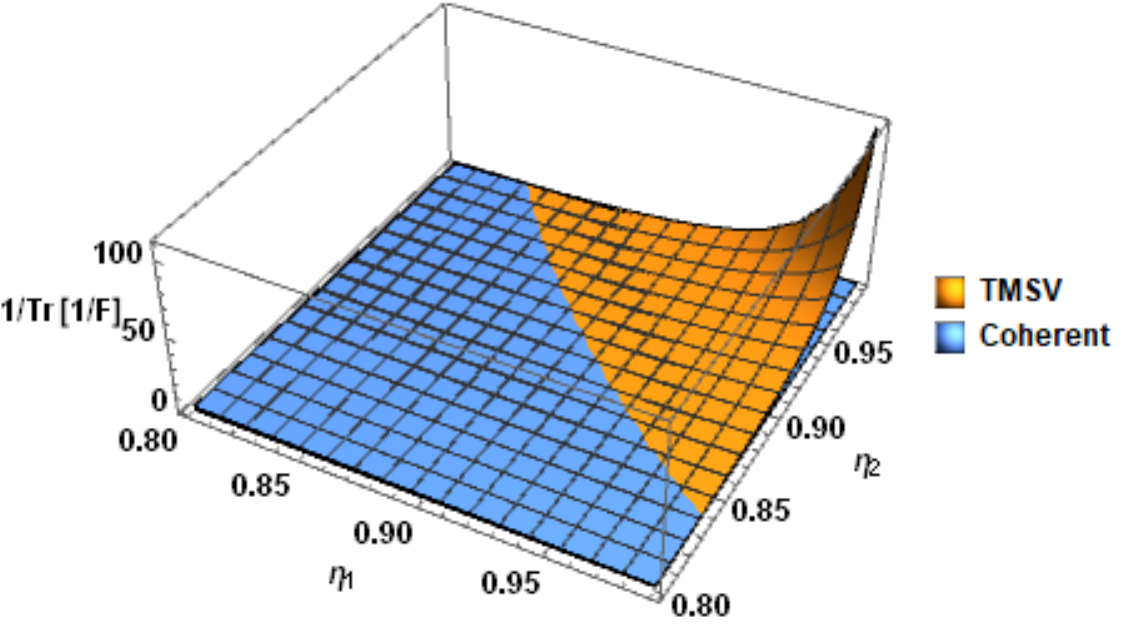}
    \caption{Sensitivity (i.e., inverse of minimum total variance) versus loss parameters $\eta_1$ and $\eta_2$ in the low-loss approximation for a squeezed state with $r=1$. While the sensitivity for TMSV is significantly better than that of coherent states when loss is small, even marginal levels of loss with transmission amplitudes $\eta_i\approx 0.9$ negate the quantum advantage.}
    \label{fig:QFI TMSV Xanadu vs coh low}
\end{figure}

We expect to see a quantum advantage in the limit of low loss, so we start by expanding $\hat{\rho}$ in terms of only the three orthonormal states $\ket{\phi_{0,0}}=\ket{\psi_{0,0}}\cosh r\sqrt{1-|f|^2}$, $\ket{\phi_{1,0}}=\ket{\psi_{1,0}}\cosh r\left(1-|f|^2\right)/|f|$, and $\ket{\phi_{0,1}}=\ket{\psi_{0,1}}\cosh r\left(1-|f|^2\right)/|f|$, which occur with probabilities
\eq{
    p_{0,0}&=\frac{1}{\cosh^2 r(1-|f|^2)}\\
    p_{1,0}&=\frac{1-\eta_1^2}{\eta_1^2}\frac{|f|^2}{\cosh^2 r(1-|f|^2)^2}\\
    p_{0,1}&=\frac{1-\eta_2^2}{\eta_2^2}\frac{|f|^2}{\cosh^2 r(1-|f|^2)^2}.
} Each of the derivatives $\partial_{\eta_i}|\phi_{j,k}\rangle$ is orthogonal to all of the states $|\phi_{j,k}\rangle$, so we only need to compute how the probabilities depend on the parameters in question.\footnote{The relative phase $\varphi$ has no further effect on the QFIM.} Expressing the QFIM in terms of the total energy $E=2\sinh^2 r$, we find, suppressing terms of order $\mathcal{O}(1-\eta_1,1-\eta_2)$,
\eq{
    \boldsymbol{\mathsf{F}}_{\mathrm{TMSV}}(\eta_1,\eta_2)\approx E\begin{pmatrix}\frac{1}{1-\eta_1}-\left(\frac{3}{2}+5E\right)&-4-3E\\
    -4-3E&\frac{1}{1-\eta_2}-\left(\frac{3}{2}+5E\right)
    \end{pmatrix}.
} The $1/(1-\eta_i)$ scaling guarantees the same advantage as Fock states relative to classical states in the small-loss limit.\footnote{One can verify the factors of 2 on the diagonals of the QFIMs: Classical states can have $4|\alpha|^2=2E$, Fock states achieve $4m/[(1+\eta_i)(1-\eta_i)]=2E/[(1+\eta_i)(1-\eta_i)]\approx E/(1-\eta_i)$, and TMSV attains $E/(1-\eta_i)$.}
This advantage can be seen in Fig. \ref{fig:QFI TMSV Xanadu vs coh low} for a particular energy. 

The QFIM is not guaranteed to be saturable for multiparameter estimation. In this scenario, saturability is possible for any measurement that reproduces the probability distribution $\{p_{0,0}, p_{1,0}, p_{0,1}\}$. A difficult-to-perform measurement saturating the QFIM could involve projecting onto the three states $|\phi_{i,j}\rangle$ mentioned above. We can also identify a readily available measurement technique: we can also reproduce this probability distribution by using PNRDs and coarse-graining the probabilities into bins corresponding to states of the form $|i\rangle_a\otimes|i\rangle_b$, $|i\rangle_a\otimes|i+1\rangle_b$, and $|i+1\rangle_a\otimes|i\rangle_b$. This optimal measurement technique from separable estimation of loss parameters is thus seen to also be optimal for the simultaneous estimation of loss parameters and intuitively follows because the two loss parameters being estimated are, in principle, compatible.

\begin{figure}
    \centering
    \includegraphics[width=\columnwidth]{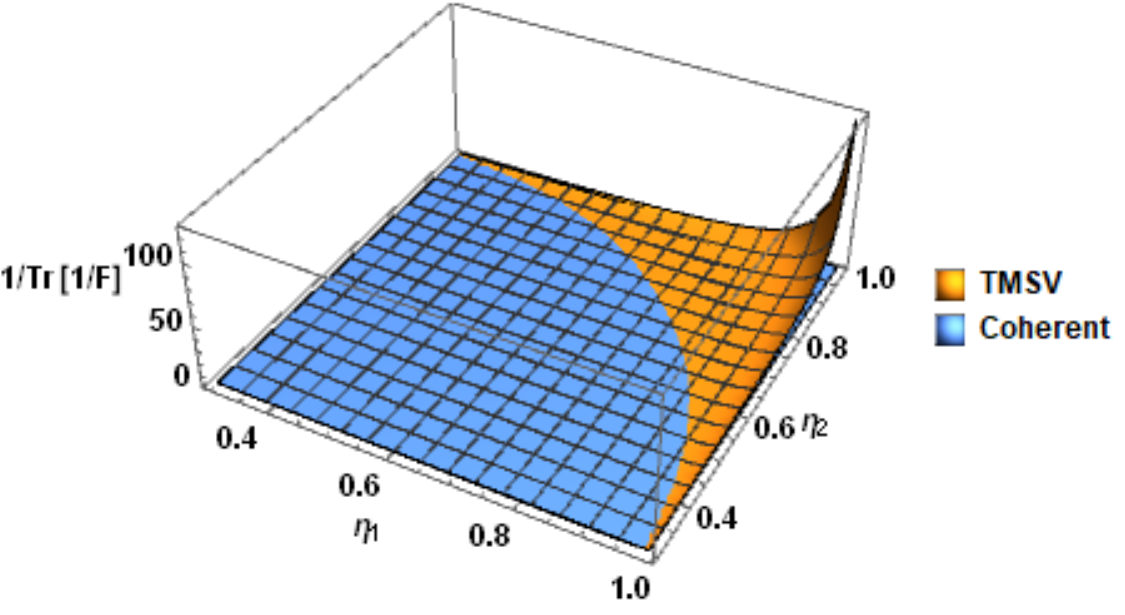}
    \caption{Sensitivity (i.e., inverse of minimum total variance) versus loss parameters $\eta_1$ and $\eta_2$ using PNRDs that can distinguish up to $10$ photons for a squeezed state with $r=1$. The sensitivity for TMSV is significantly better than that of coherent states when loss is small (Fig. \ref{fig:QFI TMSV Xanadu vs coh low}); now levels of loss with transmission amplitudes $\eta_i\approx 0.8$ negate the quantum advantage.}
    \label{fig:QFI TMSV Xanadu vs coh P10NRD}
\end{figure}

Instead of using the small-loss approximation, one can construct the photon-number probability distribution that would be detected by ideal PNRDs and calculate the Fisher information matrix from that probability distribution using the formula for the classical Fisher information $\mathsf{H}_{ij}=\sum_k p_k^{-1} \partial_i p_k \partial_j p_k$.\footnote{The ultimate limit, the QFIM, can be calculated using techniques from Gaussian quantum information as in Appendix \ref{app:Gaussian calcs}.} The Fisher information in this case is greater than the Fisher information provided in the small-loss approximation, as that approximation neglected some of the information lying in the state, thereby increasing the amounts of loss for which a quantum advantage is still possible. This is plotted in Fig. \ref{fig:QFI TMSV Xanadu vs coh P10NRD}, where an even more pronounced quantum advantage is visible relative to Fig. \ref{fig:QFI TMSV Xanadu vs coh low}. Notably, this advantage is attainable with PNRDs and PNRDs that can resolve higher photon numbers produce a larger advantage (c.f. Fig. \ref{fig:QFI TMSV Xanadu vs coh PNRD} in Appendix \ref{app:extra figs}). We also demonstrate in Appendix \ref{app:extra figs} how both TMSV and coherent states perform comparably when both modes have a large amount of loss, but coherent states outperform TMSV when one mode has a huge amount of loss and the other a small amount.

\begin{figure}
    \centering
    \includegraphics[width=\columnwidth]{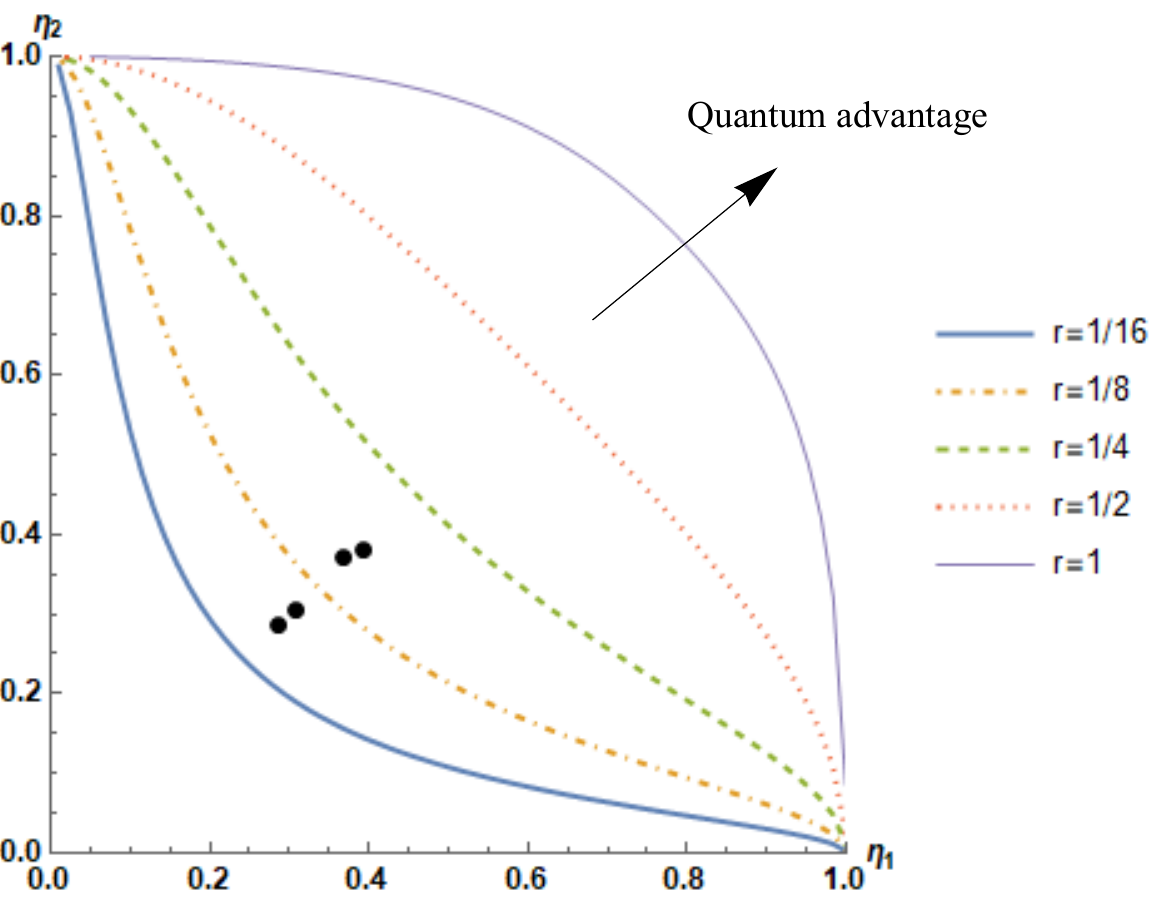}
    \caption{Crossover curves at which a TMSV with squeezing parameter $r$ outperforms a coherent state with comparable energy in the estimation of two transmission parameters $\eta_1$ and $\eta_2$, for various values of those transmission parameters. To the upper right of the crossover curves (larger transmission), TMSV provides a quantum advantage; to the lower left of the crossover curves (larger loss), coherent states outperform TMSV. Decreasing the probe energy (decreasing $r$) increases the amount of loss that can be tolerated while still providing a quantum advantage. Note that the actual transmission parameters that we measure (black dots) are all in the range from 0.2 to 0.4 such that a quantum advantage would only be noticeable for all of them with $r\lesssim 1/16$.
    }
    \label{fig:intersects}
\end{figure}

How does the quantum advantage vary with the amount of energy? We can search for the sets of loss parameters that allow for a quantum advantage in the total error versus squeezing parameter $r$, where the latter codifies the probe energy $E=2\sinh^2 r$. Inspecting Fig. \ref{fig:intersects}, we see that probes with \textit{less} energy are more forgiving for quantum enhancements: when the probe energy increases, the transmission parameters $\eta_1$ and $\eta_2$ must be larger in order to retain a quantum advantage. Quantum advantages when there is a large amount of loss are only possible when the probe state energy is low and quantum advantages when the probe has a large amount of energy are only possible when the transmission probability is large. Moreover, if we consider that probes with constant total loss or transmission are given by $\eta_1^2+\eta_2^2=\mathrm{constant}$ and form a circle centred at $\eta_1=\eta_2=0$ in Fig. \ref{fig:intersects}, we realize that quantum advantages with small $r$ are easier to obtain with balanced loss, while quantum advantages with large $r$ are easier to obtain with imbalanced loss.

\subsubsection{Three-parameter estimation}
Sequential loss channels may be combined into a single loss parameter, but the same is not true if the loss occurs before the TMSV is generated or during its generation. The TMSV is generated by a coherent state with amplitude $r$, in units normalized by the interaction strength and interaction time of the generation process, which will diminish to $\eta_0 r$ as in Fig. \ref{fig:schematic}\textbf{b)} through the usual input-output relation of Eq. \eqref{eq:input output amplitude}. It is not possible to independently distinguish the coherent state amplitude, interaction time, nonlinear interaction strength, loss, and generation efficiency from the single squeezing parameter. A realistic multiparameter loss estimation procedure must then estimate all three of $\eta_0$, $\eta_1$, and $\eta_2$ for a known $r$ or, equivalently, estimate the effective amplitude $\eta_0 r\to r$ and the two transmission amplitudes $\eta_1$ and $\eta_2$. We follow the latter procedure here.

\begin{figure}
    \centering
    \includegraphics[width=\columnwidth]{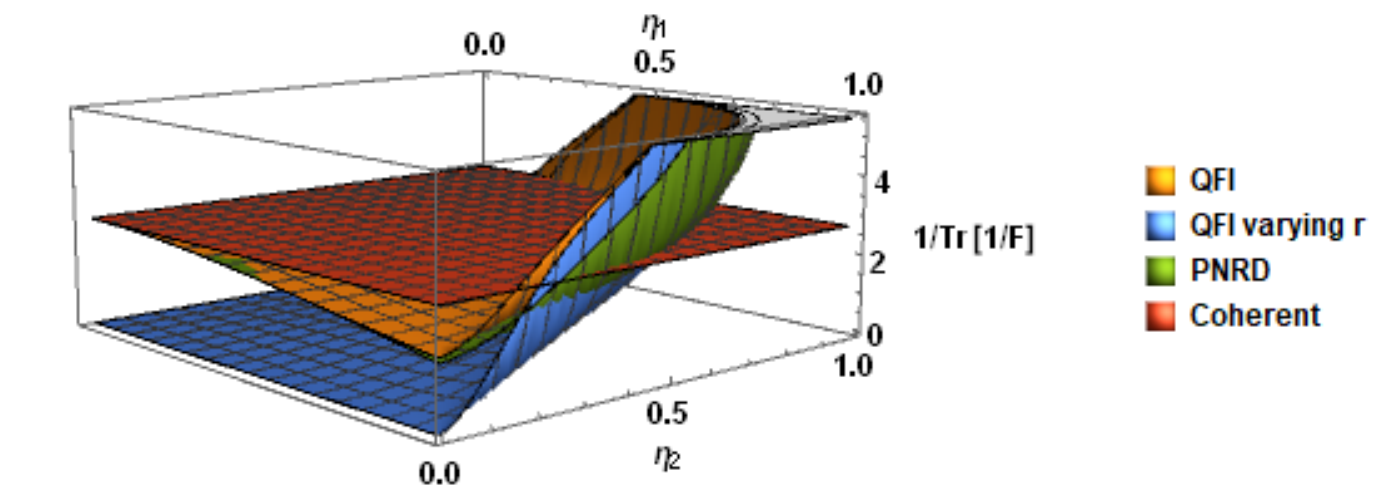}
    \caption{Sensitivity (i.e., inverse of minimum total variance) versus loss parameters as in Fig. \ref{fig:QFI TMSV Xanadu vs coh P10NRD}, now supplemented with curves for the two-parameter QFIM that behaves very similarly to the PNRD result and for the two-component subspace of the three-parameter QFIM, which conveys much less information when loss is large. We zoom in to better display the differences between the various approaches; as $\eta_1$ and $\eta_2$ grow toward 1, all of the curves other than the coherent state tend toward infinite precision as in Fig. \ref{fig:QFI TMSV Xanadu vs coh P10NRD}.}
    \label{fig:QFI TMSV Xanadu vs coh P10NRD nuisance}
\end{figure}

The QFIM for this three-parameter estimation problem is calculated in Appendix \ref{app:Gaussian calcs}, representing the ultimate limit with which the three transmission or amplitude parameters may be estimated when $r$ is unknown. We find the three simultaneous limits
\eq{
\Var \eta_1&\geq \frac{(1-\eta_1^2)(2/E+1-\eta_2^2)}{4\eta_2^2},\\
\Var \eta_2&\geq \frac{(1-\eta_2^2)(2/E+1-\eta_1^2)}{4\eta_1^2},\\
\Var r&\geq \frac{1}{2}+\frac{1-\eta_1^2-\eta_2^2}{4\eta_1^2\eta_2^2}.
\label{eq:variance bounds TMSV nuisance}
} In the large-transmission regime, the uncertainties on $\eta_1$ and $\eta_2$ match those obtained using Fock states or TMSV with PNRDs, representing the ultimate limit, and the estimates of $\eta_1$ and $\eta_2$ become uncorrelated.
In the large-loss regime, however, the introduction of a varying $r$ leads to much worse precision bounds for estimating all three parameters [also see Fig. \ref{fig:QFI TMSV Xanadu vs coh P10NRD nuisance} for comparing $(\Var \eta_1+\Var\eta_2)^{-1}$]. In the large-energy regime, the variances of $\eta_1$ and $\eta_2$ scale as $1/E$, just as for classical light [Eq. \eqref{eq:QFIM coherent}], so the crossover curves only show dependence on $r$ in Fig. \ref{fig:QFI TMSV Xanadu vs coh P10NRD nuisance} when $r$ is large. Now, \textit{how} the sum of variances $\Var \eta_1+\Var\eta_2$ is minimized for a fixed total $\eta_1^2+\eta_2^2$ does not depend on $E$ and only depends on the total transmission: when the total transmission is greater than 50\%, the sum of variances is minimized when the loss is maximally imbalanced (when one has $\eta_1=1$ or $\eta_2=1$) and, when the total transmission is less than $50\%$, the sum of variances is minimized when the loss is maximally balanced.
One final note is that the optimal estimation of $r$, which might be deemed a nuisance parameter, occurs when the average transmission probability of the two modes is exactly one half. This means that the optimal setup for estimating $\eta_1$ and $\eta_2$ will not be optimal for estimating $r$, demonstrating the qualitative difference between the loss in the generation of the TMSV and the loss after it is generated (again see Fig. \ref{fig:schematic}\textbf{b)}).

\begin{figure}
    \centering
    \includegraphics[width=\columnwidth]{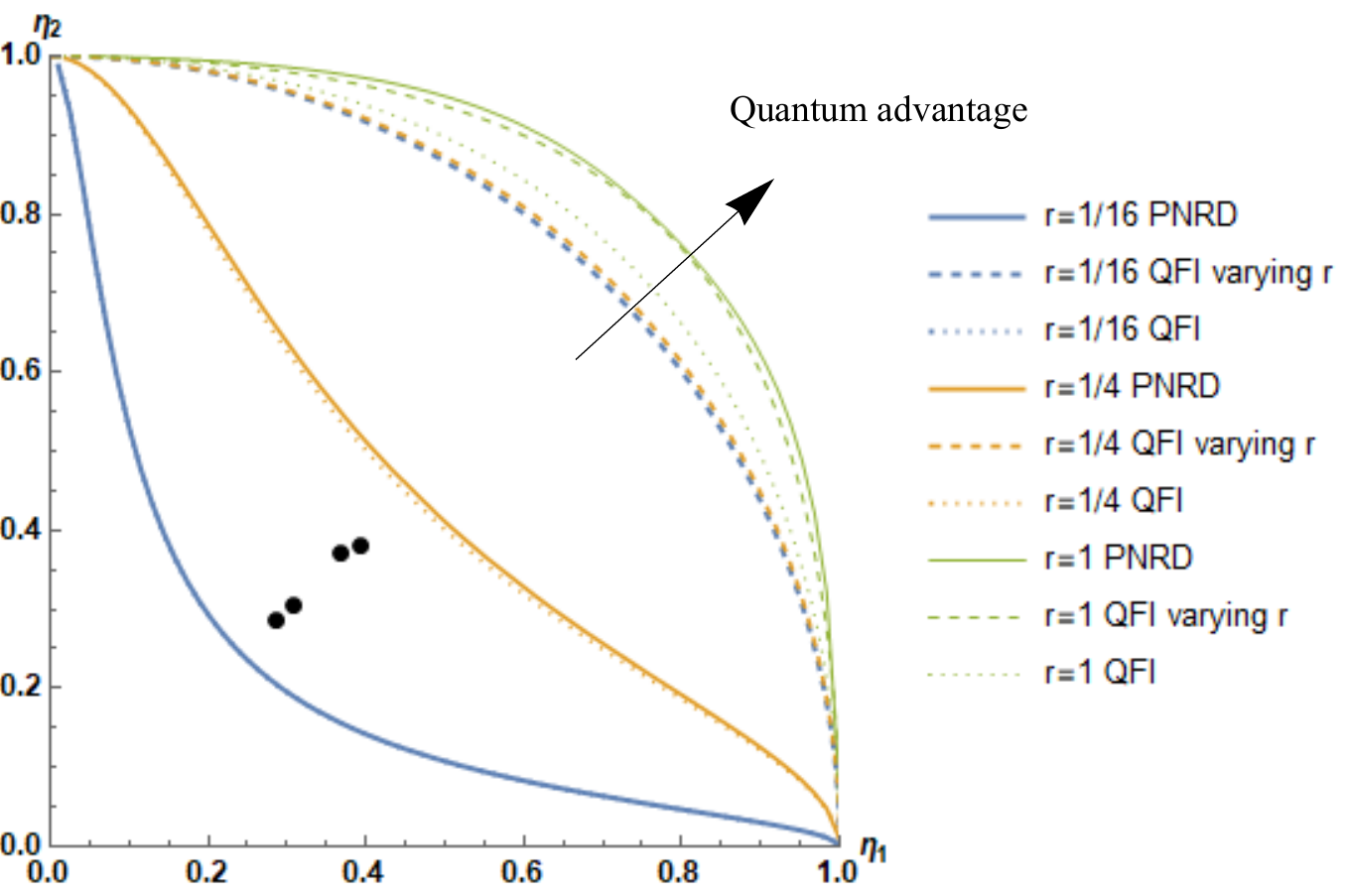}
    \caption{Crossover curves at which a TMSV with squeezing parameter $r$ outperforms a coherent state with comparable energy in the estimation of the two transmission parameters $\eta_1$ and $\eta_2$. To the upper right of the curves (larger transmission), TMSV provides a quantum advantage. The lines from Fig. \ref{fig:intersects} are plotted as solid lines, corresponding to the information gleaned about the pair of parameters when using PNRDs. The dotted lines correspond to the maximal information one could ever hope to get, which for small $r$ is approximately the same as the information from the PNRDs. The dashed lines correspond to the maximal information one could ever hope to get when one includes the effects of the ``nuisance parameter'' $r$ being able to change, which decreases the total information and therefore leads to more stringent requirements (higher transmission) for outperforming classical light. In the large energy regime, the results from using PNRDs match this ultimate bound.}
    \label{fig:intersects QFI}
\end{figure}

We can construct a plot like that of Fig. \ref{fig:intersects} using the QFIM constructed for the three-parameter estimation scheme, the QFIM for the two-parameter scheme that ignores variations in $r$, and the Fisher information from the two-parameter scheme using PNRDs (Fig. \ref{fig:intersects QFI}). While the threshold transmissivity for finding a quantum advantage is approximately the same for the PNRD scheme and the two-parameter QFIM, because the PNRD scheme is very close to the optimal one, the threshold transmissivity for the three-parameter scheme is much higher. This is because variations in $r$ dramatically increase the inverse of the QFIM due to correlations with the other two parameters of interest.

\subsubsection{Dark counts and spurious photons}
The variances given by Eq. \eqref{eq:variance bounds TMSV nuisance} are the lowest we could ever expect for input TMSV states. If the PNRDs are imperfect, however, this limit will not quite be saturated. One possible mitigating effect is when the PNRDs detect extra photons that did not originate from the TMSV. These could be stray photons from outside the experiment, pump photons that did not get filtered, or dark counts from the PNRDs themselves. 

Without spurious counts, the probability of measuring $m$ photons is given by the expectation value of the projection operator
\eq{
    \ket{m}\bra{m}=:\frac{\hat{n}^m\eu^{-\hat{n}}}{m!}:,
}where $\hat{n}=\had\ha$ is the photon-number operator and $:\cdot :$ is the normal ordering operation that moves all $\ha$ operators to the right of all $\had$ operators.
All together, spurious counts increase $\hat{n}$ by some number $\nu$, so the probability of measuring $m$ photons changes to a sum of expectation values of projectors \cite{Sperlingetal2012}
\eq{
    p_m=\expct{:\frac{\left(\hat{n}+\nu\right)^m\eu^{-\left(\hat{n}+\nu\right)}}{m!}:}=\eu^{-\nu}\sum_{k=0}^m \frac{\nu^{m-k}}{(m-k)!}\expct{:\frac{\hat{n}^k\eu^{-\hat{n}}}{k!}:},
}
effecting the transformation
\eq{
    \ket{m}\bra{m}\to \eu^{-\nu}\sum_{k=0}^m \frac{\nu^{m-k}}{(m-k)!}\ket{k}\bra{k}.
}
Each detector may have its own rate $\nu_i$ of detecting spurious photons, so we simultaneously estimate these parameters in our scheme and hope they do not diminish the precision of the other parameters too drastically.

\section{Experiment}
\subsection{Data acquisition and processing}

We accessed the X8 chip on October 14, 2021, which provides four parallel pairs of modes in which we perform our three-parameter sensing protocol, using the \texttt{RemoteEngine} class from Xanadu's Strawberry Fields library \cite{Killoranetal2019}. We created a program
to initialize four copies of the TMSV state given in Eq. \eqref{eq:TMSV} with $r=1$, which then propagated along the integrated photonic chip and through optical fibers to eight PNRDs that counted the numbers of photons arriving at the detectors at the end of the circuit (see Ref. \cite{Arrazolaetal2021} Fig. 1 for a diagram and picture of the X8 chip and Ref. \cite{Arrazolaetal2021} \textsection Methods for more details about the apparatus, including the microring resonators that generate the squeezed light and the transition-edge sensors that perform the photon-number detection). Since the loss information is encoded in the \textit{distribution} of photon numbers, we created histograms for the photon-number distributions recorded for each pair of modes after repeating the program $10^6$ times, then repeated the entire process $10^3$ times. Standard statistical techniques such as maximum likelihood estimation (MLE) could then be performed to estimate the four triads of loss parameters. The source code and recorded data can be found on GitHub at \url{https://github.com/AaronGoldberg9/transmission_TSMV_X8}. 

\begin{figure}
    \centering
    \includegraphics[width=\columnwidth]{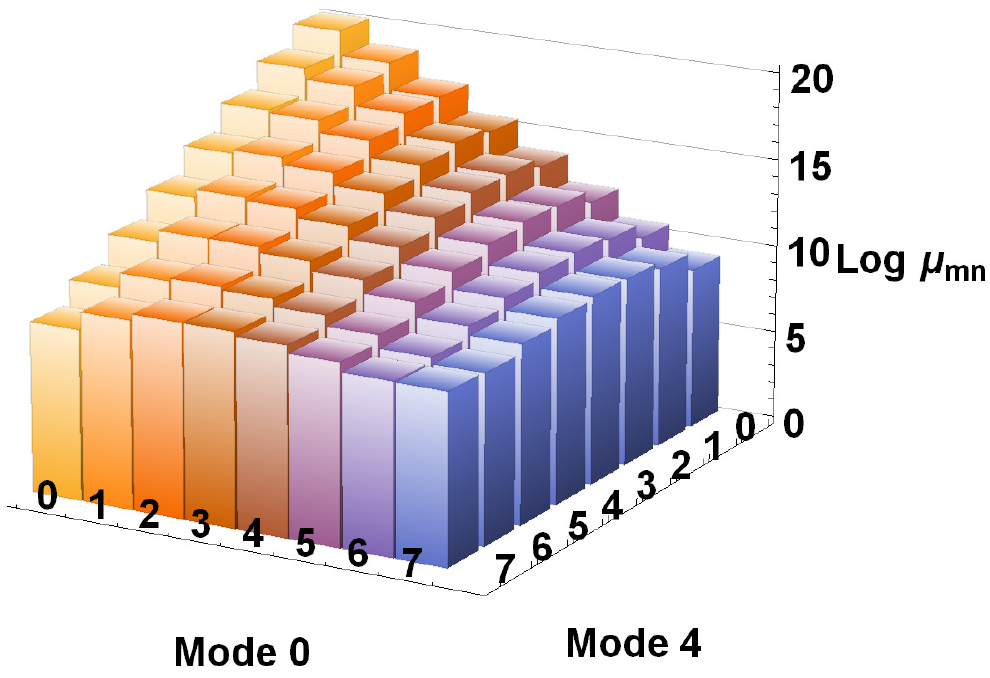}
    \caption{Semilog plot of the average measured photon-number distribution for TMSV states in modes 0 and 4 on Xanadu's X8 chip.}
    \label{fig:Log histograms 04}
\end{figure}

The average probability distributions for the first pair of modes for all $10^9$ measurements are plotted in Fig. \ref{fig:Log histograms 04} (see also Figs. \ref{fig:Log histograms 15},  \ref{fig:Log histograms 26}, and  \ref{fig:Log histograms 37} in Appendix \ref{app:extra figs} for the same plots of the other mode pairs). The average number of photons in each of the modes, labeled from 0 to 7, is found to be 0.442, 0.310, 0.417, 0.292, 0.488, 0.344, 0.429, and 0.323 which should be compared to the expected number $\sinh^2 r\approx 1.38$ in each mode. These data naively correspond to a transmission of $\approx 20-35\%$ of light in each mode, with loss being introduced anywhere from the generation of the TMSV states to the fiber coupling to the fiber propagation to the detection events. By locating these upper limits on the transmission parameters of $0.45-0.6$ on Fig. \ref{fig:intersects}, not shown, we anticipate that the amount of loss in the present setup will preclude us from reaching a quantum advantage in sensitivity. \textit{However}, setting $r=1$ using Strawberry Fields only serves to turn on the pump laser that then gets split into one pump for each of the four mode pairs; the actual pump powers lead to much larger, yet unspecified, squeezing parameters $r$, which also imply the presence of higher loss \cite{Mahler2022}.

For each of the $10^3$ trials and each pair of modes, we perform MLE to find which values of $\eta_1$, $\eta_2$, $r$ (where $r$ encompasses both the initial pump power and all losses or inefficiencies prior to TMSV generation, instead of keeping the bulky notation $\eta_0 r$), $\nu_1$, and $\nu_2$ in the conditional photon-number-probability distribution   [c.f. Eq. \eqref{eq:rho after TMSV loss}]
\eq{
    p(m,n|\eta_1,\eta_2,r,\nu_1,\nu_2)=\eu^{-\nu_1-\nu_2}\sum_{k=0}^m\sum_{l=0}^n\frac{\nu_1^{m-k}\nu_2^{n-l}}{(m-k)!(n-l)!}\\
    \times \bra{k}\otimes\bra{l}\hat{\rho}(\eta_1,\eta_2,r)\ket{k}\otimes\ket{l}
}
best reproduce the measured photon-number probability distributions that we will refer to as $q(m,n)$. In the limit of large numbers of experiments, MLE provides an unbiased estimate of the true underlying parameters with the lowest variances possible (saturating the Cram\'er-Rao lower bound). Our source code for this procedure is also available on GitHub.

What is the likelihood function required for MLE? For a discrete probability distribution, we aim to find the set of estimated parameters $\hat{\pmb{\theta}}\equiv(\hat{\eta}_1,\hat{\eta}_2,\hat{r},\hat{\nu}_1,\hat{\nu}_2)$\footnote{We use the symbol $\pmb{\theta}$ instead of $\pmb{\eta}$ to repeatedly represent the five parameters for a given pair of modes, which constitute a single set of independent parameters being estimated, in order to avoid confusion with the eight loss parameters in the eight different modes across the four parallel experiments.} that minimize the Kullback-Leibler divergence $D_{\mathrm{KL}}$ between the measured probability distribution and our parametrized probability distribution $p(m,n|\hat{\pmb{\theta}})$. This can be seen by assuming each of the $\mu=10^6$ shots to be statistically independent, with probability distribution depending on the \textit{underlying} values of the parameters $p(m,n|\pmb{\theta})$, such that the overall likelihood function is given by the product
\eq{
\mathcal{L}(\pmb{\theta}|\{m_k,n_k\}_k)=\prod_{k=1}^\mu p(m_k,n_k|\pmb{\theta}),
} where $m_k$ and $n_k$ are the number of photons recorded at the two detectors for the $k$th measurement run. Noticing that we can rearrange this likelihood product by counting the number of times each of the distinct measurement events $(m,n)$ arises, which we can refer to by $\mu_{mn}=q(m,n)\mu$, the likelihood becomes
\eq{
\mathcal{L}(\pmb{\theta}|\{\mu_{mn}\}_{mn})=\prod_{m,n} p(m,n|\pmb{\theta})^{\mu_{nm}}.
} 

We can now directly find the parameters that maximize the likelihood function, which also maximize the log-likelihood function. We write
\eq{
    \hat{\pmb{\theta}}&=\arg\max_{\pmb{\theta}}\ln\mathcal{L}(\pmb{\theta}|\{\mu_{mn}\}_{mn})\\
    &=\arg\max_{\pmb{\theta}}\sum_{m,n}\frac{\mu_{nm}}{\mu}\ln\left[p(m,n|\pmb{\theta})\right]\\
    &=\arg\min_{\pmb{\theta}}\sum_{m,n}q(m,n)\ln\left[\frac{q(m,n)}{p(m,n|\pmb{\theta})}\right]\\
    &=\arg\min_{\pmb{\theta}} D_{\mathrm{KL}}\left[\{q(m,n)\}||\{p(m,n|\pmb{\theta})\}\right],
}where we repeatedly added constant values to the quantity being optimized and reversed the extrema by inverting the arguments of the logarithms. Numerically minimizing this Kullback-Leibler divergence $D_{\mathrm{KL}}$ over all values of $\eta_1$, $\eta_2$, $r$, $\nu_1$, and $\nu_2$ directly gives us MLE's estimator for the underlying parameters. 

The one numerical caveat is that there may be events $(m,n)$ that do not arise, such that the probabilities $p(m,n)$ that contribute to $D_{\mathrm{KL}}$ do not quite sum to unity. To surmount this, we use the standard method from Refs. \cite{Hradiletal2006,Rehaceketal2008} (there attributed to Fermi under the title ``extended maximum likelihood'') of normalizing each $p(m,n)$ by the total of the probabilities for the \textit{contributing} events with $q(m,n)>0$. In the present case, the difference between the results with this normalization and without it are negligible, because the PNRDs capture the vast majority of the nonnegligible values of $p(m,n)$.

Another method for obtaining precision estimates is to combine all of the $10^9$ measurements into a single MLE procedure, ensuring that the likelihood function is gradientless at the estimated parameters, and using the concavity of the likelihood function evaluated at $\pmb{\theta}=\hat{\pmb{\theta}}$ to determine the measured/estimated Fisher information's components \cite{Rehaceketal2008} (see Appendix \ref{app:FI from likelihood} for a brief derivation):
\eq{
    \mathsf{F}_{jk}
    &=\sum_{mn}\mu_{mn}\frac{\partial \ln p(m,n|\hat{\pmb{\theta}})}{\partial \eta_j}\frac{\partial \ln p(m,n|\hat{\pmb{\theta}})}{\partial \eta_k}.
    \label{eq:FI from observed likelihood}
}
We perform this gold standard procedure and compare it to the aggregated estimates from performing MLE $10^3$ times on the respective trials.

\subsection{Results}
\begin{figure}
    \centering
    \includegraphics[width=\columnwidth]{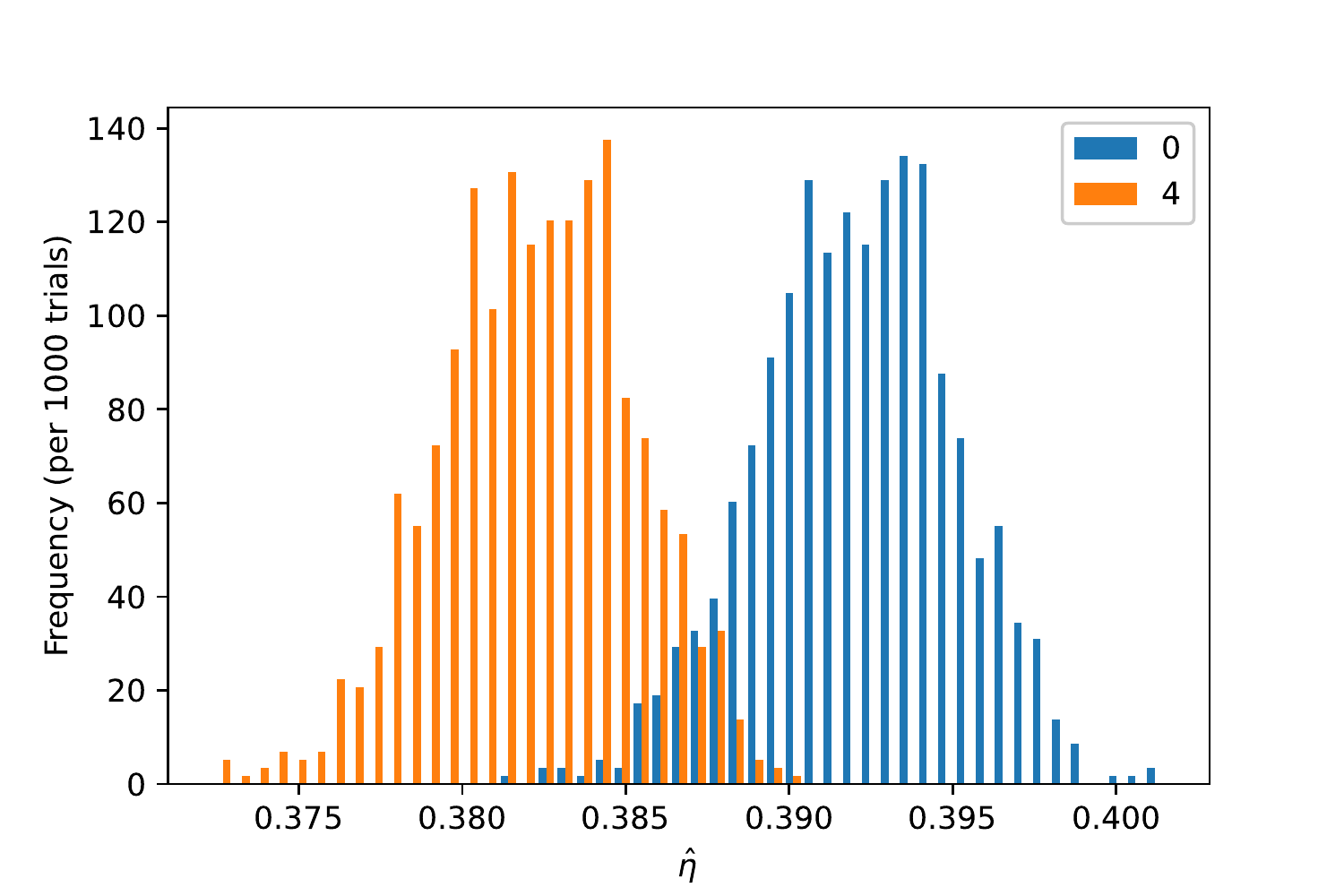}
    \caption{Estimated transmission parameters for each of the $10^3$ trials with TMSV in the pair of modes 0 and 4. The mean values of these data are 0.392(3) and 0.382(3), with the first digit of their standard deviations shown in parentheses. This corresponds to transmission probabilities of $\eta_1^2\approx 0.154$ and $\eta_2^2\approx 0.146$.}
    \label{fig:estimator hist eta 04 5par}
\end{figure}

We repeat the estimation procedure for all $M=10^3$ trials and analyze the set of estimated transmission parameters. These results
for the TMSV in modes 0 and 4 are plotted in Figs. \ref{fig:estimator hist eta 04 5par} and \ref{fig:estimator hist r 04 5par} and the rest of the distributions are displayed in Appendix \ref{app:extra figs} in Figs. \ref{fig:estimator hist eta 15 5par}, \ref{fig:estimator hist r 15 5par}, \ref{fig:estimator hist eta 26 5par}, \ref{fig:estimator hist r 26 5par},  \ref{fig:estimator hist eta 37 5par}, and \ref{fig:estimator hist r 37 5par}. The MLE estimates are approximately Gaussian for all of the parameters, qualitatively demonstrating convergence of the estimation protocol. We find that the squeezing parameters are rather large, all approximately $1.3$, leading to overall transmission probabilities of about $8\%$ to $15\%$ in each mode (first data column in Table \ref{tab:results MLE 5par}). We also find the dark count rates to range from 0.03 to 0.08 in each mode, which is much higher than one would expect from the detectors alone \cite{Milleretal2003}, so we suspect that some of the pump light is contributing to the spurious detection counts. This is confirmed by running the same circuit with the pump turned off, which reduces the dark count rates to $0.01$, 0.01, 0.03, 0.02, 0.02, 0.02, 0.03, and 0.04.
\begin{figure}
    \centering
    \includegraphics[width=\columnwidth]{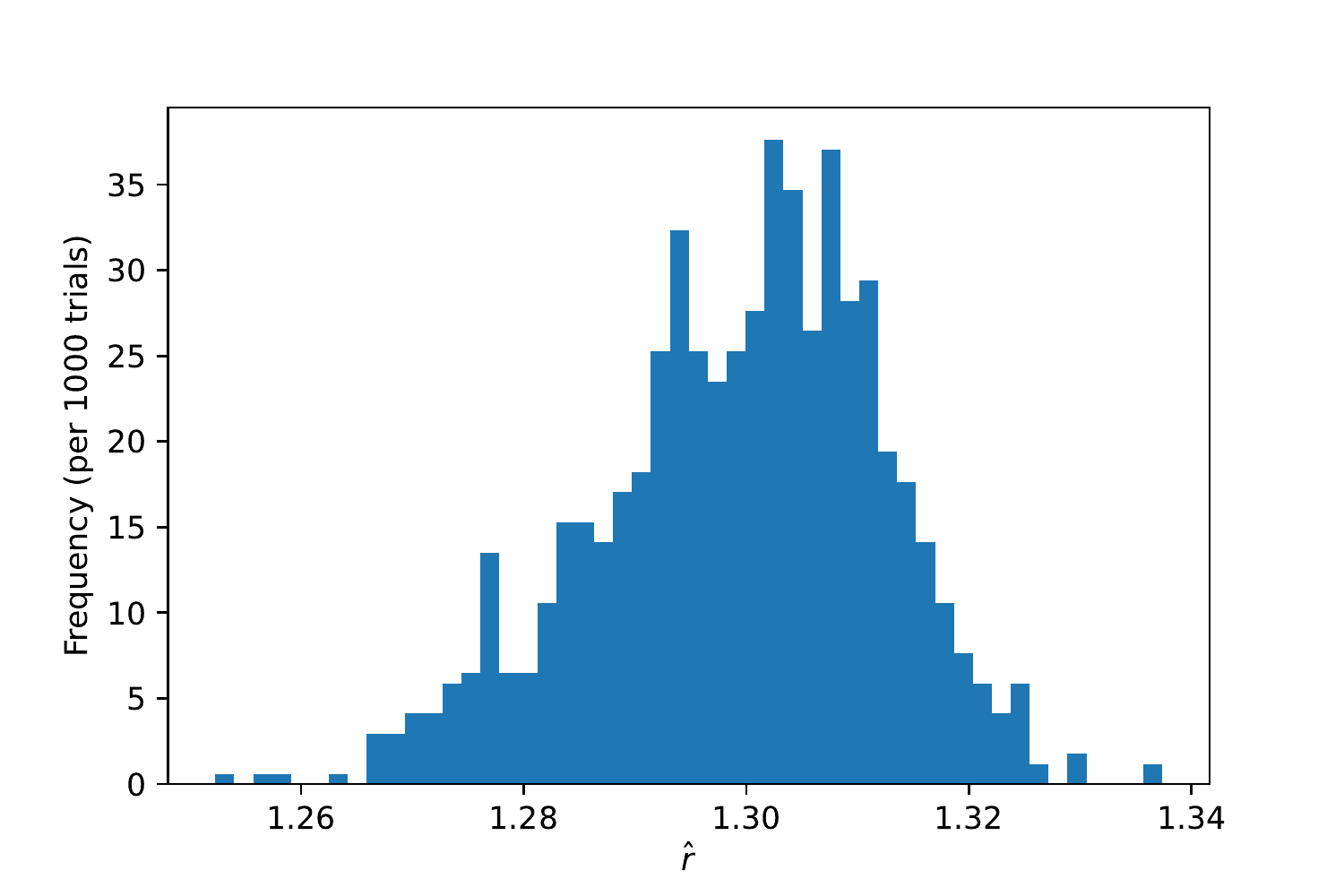}
    \caption{Estimated effective squeezing parameter for each of the $10^3$ trials with TMSV in the pair of modes 0 and 4. The mean value of these data is  1.30(1), with the first digit of its standard deviation shown in parentheses. This corresponds to an average photon number in the generated TMSV state of $2\sinh^2 r\approx 5.77$.}
    \label{fig:estimator hist r 04 5par}
\end{figure}

Can we confirm that the estimators have converged to the ``true'' values of the underlying parameters? From the histograms, one can appreciate that the results are approximately normally distributed.
A quick test for the convergence properties is to use the equivariance property of MLE: we perform MLE using a different parametrization, such as using the transmission probabilities $q_i=\eta_i^2$, and inspect whether $\sqrt{\expct{\hat{q}_i}}=\expct{\hat{\eta}_i}$, where the expectation value is taken over all MLE trials. We compare all of these data in Table \ref{tab:results MLE 5par}; near-identical results are visible with all other reparametrizations that we verified, including the aforementioned $\phi=\arccos\eta$ parameter. In Table \ref{tab:results MLE 5par}, we also include the results for performing the MLE on the aggregate data from all $10^9$ trials, whose covariances are determined by the concavity of the likelihood function at the estimated parameter values through Eq. \eqref{eq:FI from observed likelihood}.
This significant agreement with the means and standard deviations for our estimators $\hat{\eta}_i$ as quoted in the abstract confirms the convergence of our estimators. The most significant outlier from ideal convergence comes from the mode pair 2 and 6 with only a 1\% deviation between methods, with all of the rest of the means agreeing to even higher precision.

\begin{table*}[]
\centering
\caption{Estimated transmission parameters and their variances for each of the mode pairs, using different MLE methods. Highlighted in yellow are the estimates from the aggregated photon-number statistics from all $10^9$ shots that differ in the second significant digit from the average estimates over the $10^3$ trials with $\mu=10^6$ shots each. Highlighted in red are the variances of the MLE estimates from the $10^3$ trials that are smaller than the ultimate limit achievable with either a coherent state $|\alpha\rangle$ or a Fock state $|m\rangle\otimes|m\rangle$ with comparable energy.}
\label{tab:results MLE 5par}
\begin{tabular}{|c|c|c|c|c|c|c|c|c|c|c|c|}
\hline
\begin{tabular}[c]{@{}c@{}}Mode\\ pair\end{tabular} &
  Param. &
  \begin{tabular}[c]{@{}c@{}}Mean\\ $10^3$ trials\end{tabular} &
  $\sqrt{\expct{\hat{\theta}_i^2}}$ &
  \begin{tabular}[c]{@{}c@{}}Estimate\\ $10^9$ shots\end{tabular} &
  \begin{tabular}[c]{@{}c@{}}Var.\\ $10^3$ trials\end{tabular} &
  \begin{tabular}[c]{@{}c@{}}Var. $|\alpha\rangle$\\ $M=10^3$\end{tabular} &
  \begin{tabular}[c]{@{}c@{}}Var. $|m\rangle$\\ $M=10^3$\end{tabular} &
  \begin{tabular}[c]{@{}c@{}}Var.\\ $10^9$ shots\end{tabular} &
  \begin{tabular}[c]{@{}c@{}}Var. $|\alpha\rangle$\\ $M=10^9$\end{tabular} &
  \begin{tabular}[c]{@{}c@{}}Var. $|m\rangle$\\ $M=10^9$\end{tabular} &
  \begin{tabular}[c]{@{}c@{}}Var. QFIM\\ $M=10^9$\end{tabular} \\ \hline
 &
  $\hat{\eta}_1$ &
  0.392 &
  0.392 &
  0.39202 &
  \cellcolor[HTML]{FFCCC9}$9.1\times 10^{-6}$ &
   &
  $1.5\times 10^{-4}$ &
  $4.0\times 10^{-9}$ &
   &
  $1.5\times 10^{-10}$ &
  $1.7\times 10^{-9}$ \\ \cline{2-6} \cline{8-9} \cline{11-12} 
 &
  $\hat{\eta}_2$ &
  0.382 &
  0.382 &
  0.38206 &
  \cellcolor[HTML]{FFCCC9}$8.7\times 10^{-6}$ &
  \multirow{-2}{*}{$1.8\times 10^{-4}$} &
  $1.5\times 10^{-4}$ &
  $3.8\times 10^{-9}$ &
  \multirow{-2}{*}{$1.8\times 10^{-10}$} &
  $1.5\times 10^{-10}$ &
  $1.7\times 10^{-9}$ \\ \cline{2-12} 
 &
  $\hat{r}$ &
  1.30 &
  1.30 &
  1.3000 &
  $1.6\times 10^{-4}$ &
   &
   &
  $2.6\times 10^{-8}$ &
   &
   &
  $8.3\times 10^{-9}$ \\ \cline{2-12} 
 &
  $\hat{\nu}_1$ &
  0.034 &
  0.034 &
  0.03419 &
  $1.5\times 10^{-6}$ &
   &
   &
  $6.6\times 10^{-9}$ &
   &
   &
   \\ \cline{2-12} 
\multirow{-5}{*}{0 and 4} &
  $\hat{\nu}_2$ &
  0.066 &
  0.066 &
  0.06568 &
  $1.4\times 10^{-5}$ &
   &
   &
  $5.3\times 10^{-9}$ &
   &
   &
   \\ \hline
 &
  $\hat{\eta}_1$ &
  0.307 &
  0.307 &
  0.30706 &
  \cellcolor[HTML]{FFCCC9}$3.0\times 10^{-5}$ &
   &
  $1.6\times 10^{-4}$ &
  $6.8\times 10^{-9}$ &
   &
  $1.6\times 10^{-10}$ &
  $3.0\times 10^{-9}$ \\ \cline{2-6} \cline{8-9} \cline{11-12} 
 &
  $\hat{\eta}_2$ &
  0.305 &
  0.305 &
  0.30441 &
  \cellcolor[HTML]{FFCCC9}$1.9\times 10^{-5}$ &
  \multirow{-2}{*}{$1.8\times 10^{-4}$} &
  $1.6\times 10^{-4}$ &
  $6.6\times 10^{-9}$ &
  \multirow{-2}{*}{$1.8\times 10^{-10}$} &
  $1.6\times 10^{-10}$ &
  $3.0\times 10^{-9}$ \\ \cline{2-12} 
 &
  $\hat{r}$ &
  1.32 &
  1.32 &
  1.3238 &
  $2.7\times 10^{-4}$ &
   &
   &
  $7.2\times 10^{-8}$ &
   &
   &
  $2.4\times 10^{-8}$ \\ \cline{2-12} 
 &
  $\hat{\nu}_1$ &
  0.041 &
  0.041 &
  0.04066 &
  $3.0\times 10^{-6}$ &
   &
   &
  $5.3\times 10^{-9}$ &
   &
   &
   \\ \cline{2-12} 
\multirow{-5}{*}{1 and 5} &
  $\hat{\nu}_2$ &
  0.060 &
  0.060 &
  0.05945 &
  $4.4\times 10^{-6}$ &
   &
   &
  $4.9\times 10^{-9}$ &
   &
   &
   \\ \hline
 &
  $\hat{\eta}_1$ &
  0.369 &
  0.369 &
  0.36837 &
  \cellcolor[HTML]{FFCCC9}$1.0\times 10^{-5}$ &
   &
  $1.6\times 10^{-4}$ &
  $4.2\times 10^{-9}$ &
   &
  $1.6\times 10^{-10}$ &
  $1.9\times 10^{-9}$ \\ \cline{2-6} \cline{8-9} \cline{11-12} 
 &
  $\hat{\eta}_2$ &
  0.373 &
  0.373 &
  0.37229 &
  \cellcolor[HTML]{FFCCC9}$1.1\times 10^{-5}$ &
  \multirow{-2}{*}{$1.8\times 10^{-4}$} &
  $1.6\times 10^{-4}$ &
  $4.0\times 10^{-9}$ &
  \multirow{-2}{*}{$1.8\times 10^{-10}$} &
  $1.6\times 10^{-10}$ &
  $2.0\times 10^{-9}$ \\ \cline{2-12} 
 &
  $\hat{r}$ &
  1.26 &
  1.26 &
  1.2666 &
  $9.0\times 10^{-4}$ &
   &
   &
  $3.0\times 10^{-8}$ &
   &
   &
  $1.0\times 10^{-8}$ \\ \cline{2-12} 
 &
  $\hat{\nu}_1$ &
  0.054 &
  0.054 &
  \cellcolor[HTML]{FFFC9E}0.05311 &
  $3.1\times 10^{-6}$ &
   &
   &
  $6.4\times 10^{-9}$ &
   &
   &
   \\ \cline{2-12} 
\multirow{-5}{*}{2 and 6} &
  $\hat{\nu}_2$ &
  0.073 &
  0.073 &
  \cellcolor[HTML]{FFFC9E}0.07195 &
  $4.2\times 10^{-6}$ &
   &
   &
  $6.7\times 10^{-9}$ &
   &
   &
   \\ \hline
 &
  $\hat{\eta}_1$ &
  0.288 &
  0.288 &
  0.28730 &
  \cellcolor[HTML]{FFCCC9}$6.6\times 10^{-5}$ &
   &
  $1.7\times 10^{-4}$ &
  $7.6\times 10^{-9}$ &
   &
  $1.7\times 10^{-10}$ &
  $3.5\times 10^{-9}$ \\ \cline{2-6} \cline{8-9} \cline{11-12} 
 &
  $\hat{\eta}_2$ &
  0.287 &
  0.287 &
  0.28621 &
  \cellcolor[HTML]{FFCCC9}$8.0\times 10^{-5}$ &
  \multirow{-2}{*}{$1.8\times 10^{-4}$} &
  $1.7\times 10^{-4}$ &
  $7.2\times 10^{-9}$ &
  \multirow{-2}{*}{$1.8\times 10^{-10}$} &
  $1.7\times 10^{-10}$ &
  $3.4\times 10^{-9}$ \\ \cline{2-12} 
 &
  $\hat{r}$ &
  1.34 &
  1.34 &
  1.3425 &
  $1.5\times 10^{-3}$ &
   &
   &
  $9.0\times 10^{-8}$ &
   &
   &
  $3.1\times 10^{-8}$ \\ \cline{2-12} 
 &
  $\hat{\nu}_1$ &
  0.036 &
  0.036 &
  0.03552 &
  $1.9\times 10^{-6}$ &
   &
   &
  $4.7\times 10^{-9}$ &
   &
   &
   \\ \cline{2-12} 
\multirow{-5}{*}{3 and 7} &
  $\hat{\nu}_2$ &
  0.078 &
  0.078 &
  0.07791 &
  $1.6\times 10^{-5}$ &
   &
   &
  $4.3\times 10^{-9}$ &
   &
   &
   \\ \hline
\end{tabular}
\end{table*}

How do these results compare to the ultimate limit? Taking the estimated parameters to represent the true underlying ones, we can inspect the lowest variances that might be obtained using coherent and Fock states with the same total input energies $E=2\sinh^2 \hat{r}$. We use Eqs. \eqref{eq:QFIM coherent} and \eqref{eq:QFIM Fock} with $|\alpha|^2=|\beta|^2=E/2$ and $m=n=E/2$, respectively to find the $M=1$ results ranging from approximately 0.15 to 0.17. If we now increase to $M=10^3$ to represent the different trials performed, we see the variances in our estimates of $\eta_1$ and $\eta_2$ to differ by 1-2 orders of magnitude with respect to the limits imposed by coherent and Fock states. Remarkably, the variances obtained are \textit{smaller than the ultimate limit} provided by Fock states with comparable energies (Table \ref{tab:results MLE 5par}), implying that more care must be taken in these statistics.

If we instead increase $M$ to $10^9$ to account for the total number of experiments performed, the variances of the $10^3$ estimates easily obey the quantum Cram\'er-Rao lower bound. We can then use the techniques of Refs. \cite{Hradiletal2006,Rehaceketal2008} to analyze all of the data in aggregate, discarding the trial identifications. We then find the estimated parameters quoted in the abstract and variances closer to but always greater than the variances one would obtain from coherent and Fock states with $M=10^9$ repetitions  (Table \ref{tab:results MLE 5par}). 

One can finally compare the variances obtained thus with the ultimate limit one could ever expect to obtain \textit{using TMSV states} that take into account the possible variations in $r$, as in Eq. \eqref{eq:variance bounds TMSV nuisance}. Then, the measured variances are approximately twice to thrice the ultimate bound: in Table \ref{tab:results MLE 5par}, we observe that the estimates approximately saturate the qCRB, with variances coming within a factor of three of the ultimate quantum limit using TMSV as a probe state for simultaneously estimating all three parameters, even in the presence of nonnegligible dark counts that deteriorate the measurement.\footnote{The ``factor of two'' phrase in our abstract for the uncertainties is really $\lesssim\sqrt{3}$.} This means that they could certainly beat classical bounds if the transmission parameters were larger and is why we refer to ``at the quantum limit'' in our title.

Overall, how well do our estimates account for the underlying data? We can measure the root-mean-square (RMS) error between the probability distributions that we measured and the probability distributions that we can construct using our estimators. We find the overall RMS values 0.003 for each of the four joint probability distributions for the four pairs of modes, so the measured histograms are certainly close to what one would expect from our loss model.

\subsection{Further analysis}
\subsubsection{Arranging shots and trials}
We have estimated the uncertainty in our estimates using two methods: computing the covariance of our $10^3$ independent estimates from the $10^6$-shot experiments and using Eq. \eqref{eq:FI from observed likelihood} to estimate the covariance after aggregating all $10^9$ shots. There are many ways of interpolating between these results.

\begin{figure}
    \centering
    \includegraphics[width=\columnwidth]{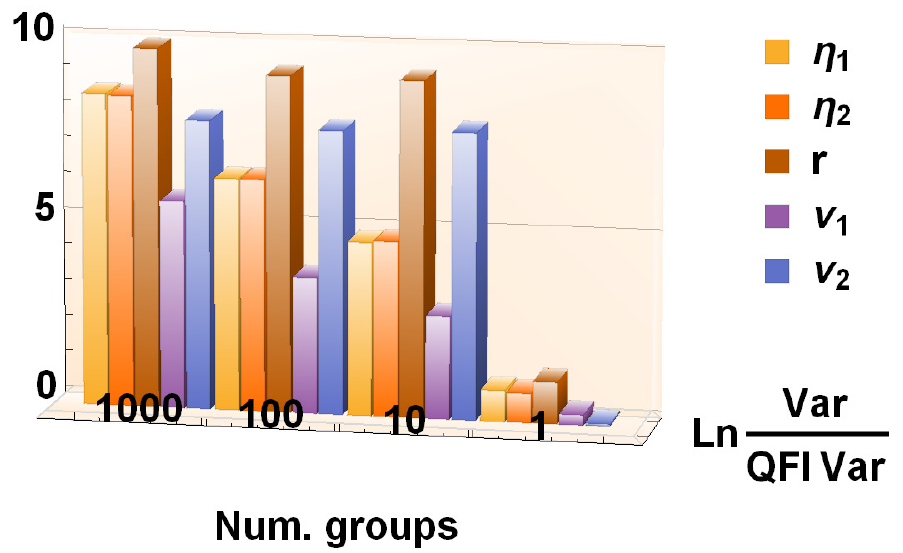}
    \caption{Variances of the MLE estimates from the data aggregated into different numbers of groups, relative to the theoretical limit of what can be obtained using TMSV, for modes 0 and 4. The variances of $\hat{\eta}_1$ and $\hat{\eta}_2$ diminish exponentially with increasing group size (decreasing number of groups), while the variances of the other estimated parameters decrease more slowly, demonstrating that more shots within a trial is superior to having more total trials with fewer shots. The QFI value for the variances of the dark count parameters $\hat{\nu}_i$ is artificially set to $10^{-9}$ for scaling purposes.}
    \label{fig:var change size 04 5par}
\end{figure}

One strategy is to combine all of the shots from some trials into a single larger trial, perform MLE on the larger trial, and repeat for the remaining trials. This tends to reduce the variances in the MLE estimates, even though fewer trials are being performed, demonstrating that it is more important to have better statistics in each trial than to have more trials. For some mode pairs, this decrease is exponential in the variances of $\hat{\eta}_1$ and $\hat{\eta}_2$, while for others the decrease is much slower (see Fig. \ref{fig:var change size 04 5par} for modes 0 and 4 and Appendix \ref{app:extra figs} Figs. \ref{fig:var change size 15 5par}, \ref{fig:var change size 26 5par}, and \ref{fig:var change size 37 5par} for the other mode pairs with less dramatic results). In fact, the opposite limit performs extremely poorly. If one considers each of the $10^9$ shots to be a single experiment, calculates what the estimated parameter values would be from each possible PNRD result, then performs statistics on these $10^9$ estimates, one obtain results not much better than a uniform distribution over all possible parameter values (the variances for each $\hat{\eta}_i$, which for a uniform distribution would be $1/12\approx 0.08$, are all between 0.07 and 0.1). These results show that $M=10^3$ trials is sufficient for convergence in all of the situations, while smaller $M$ values are already sufficient for some of the mode pairs.

Another strategy to reducing variances is through bootstrapping. We can take random sets of shots from all of our data, perform MLE on them, then repeat the process to collect statistics. If we sample without replacement, we get approximately the same results as before, still dependent on the size of the random sets. If we sample with replacement, we can arbitrarily decrease the variances in and increase the Gaussianity of our distribution of estimates, simply because we artificially increase the effective number of shots without performing more measurements. Similarly, we can perform parametric bootstrapping, where we take our estimated parameters from the aggregated data to construct a probability distribution from which we randomly generate more data sets. These methods cannot be relied upon to claim quantum advantages, however, because with them one could obtain arbitrary precisions from a single measurement.

\subsubsection{Validity of observed Fisher information}
\begin{figure}
    \centering
    \includegraphics[width=\columnwidth]{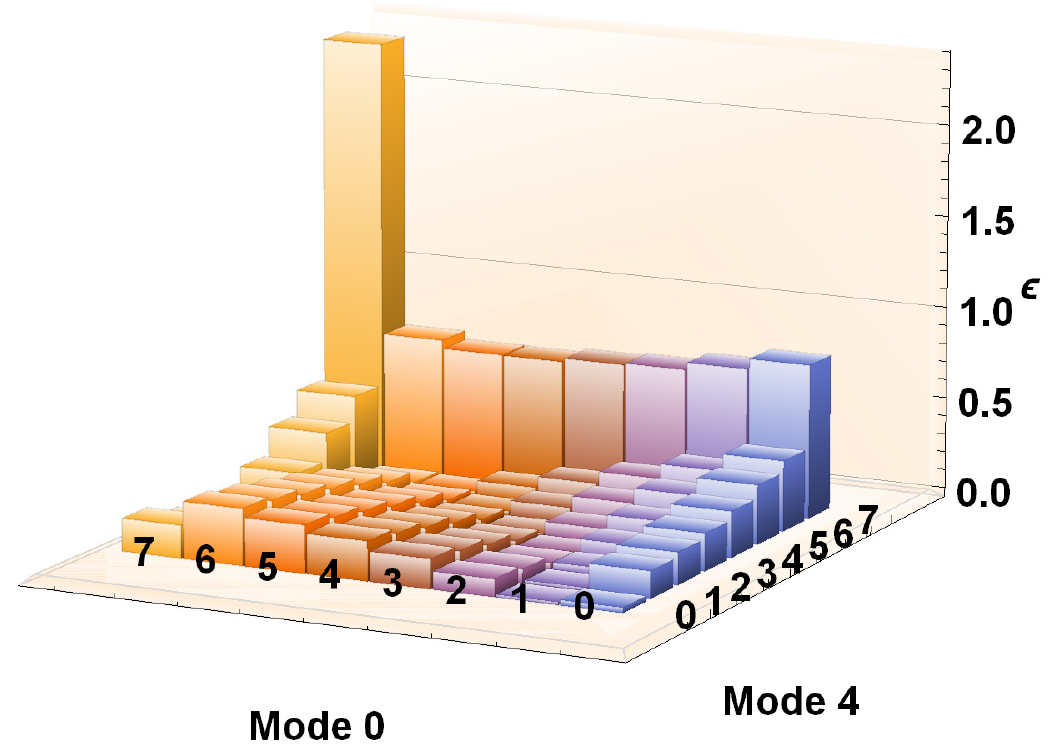}
    \caption{Absolute relative errors between the observed frequencies and expected probabilities for each photon-number-detection event in modes 0 and 4. These relative errors are sufficiently small to justify using Eq. \eqref{eq:FI from observed likelihood} to estimate the covariances of the underlying parameters. The average value is 0.24588. Note that the order of the photon numbers is opposite relative to Fig. \ref{fig:Log histograms 04} to make the data more visible.}
    \label{fig:rel err04 5par}
\end{figure}

There is an assumption used to estimate the covariance from a single aggregated trial with all of the measurement data, as in Eq. \eqref{eq:FI from observed likelihood}. Ideally, enough measurements must be performed such that $\mu_{mn}\approx \mu p(m,n|\hat{\pmb{\theta}})$; equivalently, the measured frequencies should match the probabilities that would be obtained should the estimated parameters be the true parameters. We have quoted the small RMS values between these probability distributions as $0.003$ for all of the mode pairs, which seem to imply that the observed frequencies are equally representative of the underlying probabilities for each of the mode pairs. However, we have to recall from Eq. \eqref{eq:rho after TMSV loss} and Fig. \ref{fig:Log histograms 04} that the distributions are strongly peaked at $(m,n)=(0,0)$ and that it is rare to detect a large number of photons. This means that, even though the probability distributions are well approximated on average, there may still be terms with $\mu_{mn}\not\approx \mu p(m,n|\hat{\pmb{\theta}})$ at large $(m,n)$ because of the scarcity of statistics in that regime, which precludes the use of Eq. \eqref{eq:FI from observed likelihood} for estimating variances for those experiments. To verify our assumptions, we plot in Fig. \ref{fig:rel err04 5par} the relative error parameter
\eq{
    \epsilon(m,n)=\left|\frac{\mu_{mn}/\mu-p(m,n|\hat{\pmb{\theta}})}{p(m,n|\hat{\pmb{\theta}})}\right|
} for modes 0 and 4 after aggregating all of the data into a single trial ($M=1$) with $\mu=10^9$ shots. Even though $\epsilon$ grows with increasing $m$ and $n$, it remains small enough for us to trust the estimate, with average relative error of $0.25$. In fact, $\epsilon$ is only sizeable at the limits of the largest $m$ and $n$ values, which contribute the least to the observed Fisher information. Similar plots for the other modes can be found in Appendix \ref{app:extra figs} Figs. \ref{fig:rel err15 5par}, \ref{fig:rel err26 5par}, and \ref{fig:rel err37 5par}, validating the use of Eq. \eqref{eq:FI from observed likelihood} for determining how close our results are to the ultimate limit given in Eq. \eqref{eq:variance bounds TMSV nuisance}.

\subsubsection{$10^3$ independent experiments}
Finally, we can inspect each of the $10^3$ trials individually and compute the estimated covariances for each of their measured parameters using Eq. \eqref{eq:FI from observed likelihood}. This allows us to consider each trial as measuring a different set of underlying parameters, which we found to be distributed as in Fig. \ref{fig:estimator hist eta 04 5par} and similar plots. For each of those underlying parameters, we can then compare their estimated covariances to the ultimate limit possible with TMSV probe states that is given in Eq. \eqref{eq:variance bounds TMSV nuisance}. 

Recall from Table \ref{tab:results MLE 5par} that the ratios of the observed variances to the minimum variances possible from the QFI for parameters $\hat{\eta}_1$, $\hat{\eta}_2$, and $\hat{r}$ are approximately $(2.4, 2.2, 3.1)$, $(2.3, 2.2, 3.0)$, $(2.2, 2.0, 3.0)$, and $(2.2, 2.1, 2.9)$, for mode pairs 0 and 4, 1 and 5, 2 and 6, and 3 and 7, respectively. Those results were obtained by combining all $10^9$ shots into a single effective trial. Now, when we use only $10^6$ shots per trial, we observe that every trial \textit{individually} performs nearly as well as the aggregated trial. For example, the very first trial yields variance ratios of approximately $(2.2, 2.5, 3.2)$, $(2.1, 2.3, 3.1)$, $(2.1, 2.2, 3.0)$, and $(2.0, 2.2, 2.9)$ for the respective mode pairs. Some of the uncertainties (i.e., the square roots of the variances) are better than those of the aggregated trial, some are worse, and they are all within 10$\%$ of each other. 

Similar variance ratios between the estimated parameters and the ultimate quantum limit are observed for each trial with its $10^6$ shots. We immediately conclude that a single trial could be used to well estimate the underlying parameters, making quantum-enhanced transmission or loss estimation extremely feasible on cloud quantum computers. Repeating these trials helps characterize the stability of the devices; once this stability is established, there is little extra benefit conferred in spending more time measuring the same parameters. It thus takes only \textit{seconds} to demonstrate a quantum advantage on Xanadu's X8 chip, accessible through the cloud.

\section{Conclusions}
We have performed a multiparameter loss estimation procedure on a cloud quantum computer.
Specifically, we developed the theory and practice for simultaneously estimating transmission parameters in \textit{both} branches of a two-mode squeezed-vacuum state and in its generation process, as well as the spurious count rates at each detector.
We found the amounts of loss to be too high such that they preclude a quantum enhancement relative to classical states for the given probe state energy, but that we can reach the ultimate limit for TMSV input states within a factor of two in uncertainty, such that we could observe a significant quantum enhancement if the amount of loss was lower. We found that increasing the number of photons that a detector can resolve increases the range of parameters for which one can see a quantum enhancement. We also analyzed different methods of aggregating measurement results for performing optimal estimation procedures and provided insight into when and how to trust results from maximum likelihood estimation. These showcase the readiness of available quantum technologies for demonstrating enhances in significant metrological tasks.

\begin{acknowledgments}
    The authors acknowledge that the NRC headquarters is located on the traditional unceded territory of the Algonquin Anishinaabe and Mohawk people. The authors also thank Xanadu for giving them early access to the X8 device and acknowledge support from NRC's Quantum Sensing Challenge program. AZG acknowledges funding from the NSERC PDF program.
\end{acknowledgments}

\clearpage
\appendix

\section{Extra figures}
\label{app:extra figs}
Here we collate supplementary figures that were mentioned in the main text, in order of their appearance.

\begin{figure}[h]
    \centering
    \includegraphics[width=\columnwidth]{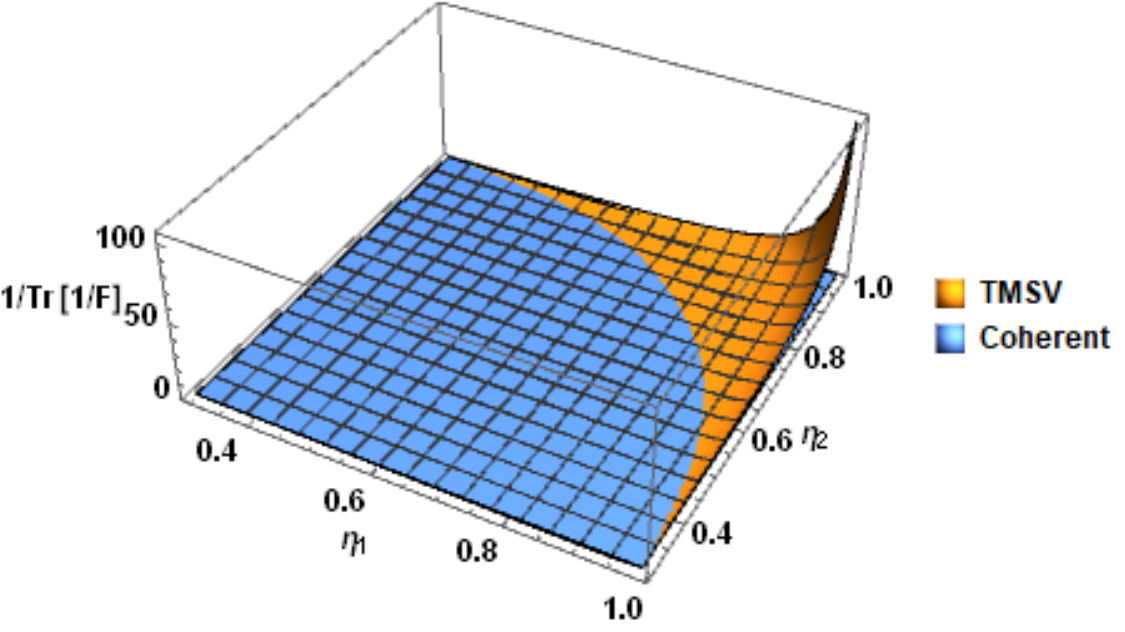}
    \caption{Sensitivity (i.e., inverse of minimum total variance) versus loss parameters $\eta_1$ and $\eta_2$ using PNRDs that can distinguish up to $5$ photons for a squeezed state with $r=1$. The sensitivity for TMSV is significantly better than that of coherent states when loss is small, now levels of loss with transmission amplitudes $\eta_i\approx 0.8$ negate the quantum advantage.}
    \label{fig:QFI TMSV Xanadu vs coh PNRD}
\end{figure}

We first demonstrate the differences in sensitivity on can achieve by using PNRDs capable of resolving larger photon numbers. 
Using PNRDs that can resolve up to 5 photons yields a larger quantum advantage (Fig. \ref{fig:QFI TMSV Xanadu vs coh PNRD}) than the approximate method in Fig. \ref{fig:QFI TMSV Xanadu vs coh low}, whereas using PNRDs that can resolve up to 10 photons outperforms both, as in Fig. \ref{fig:QFI TMSV Xanadu vs coh P10NRD}.

\begin{figure}
    \centering
    \includegraphics[width=\columnwidth]{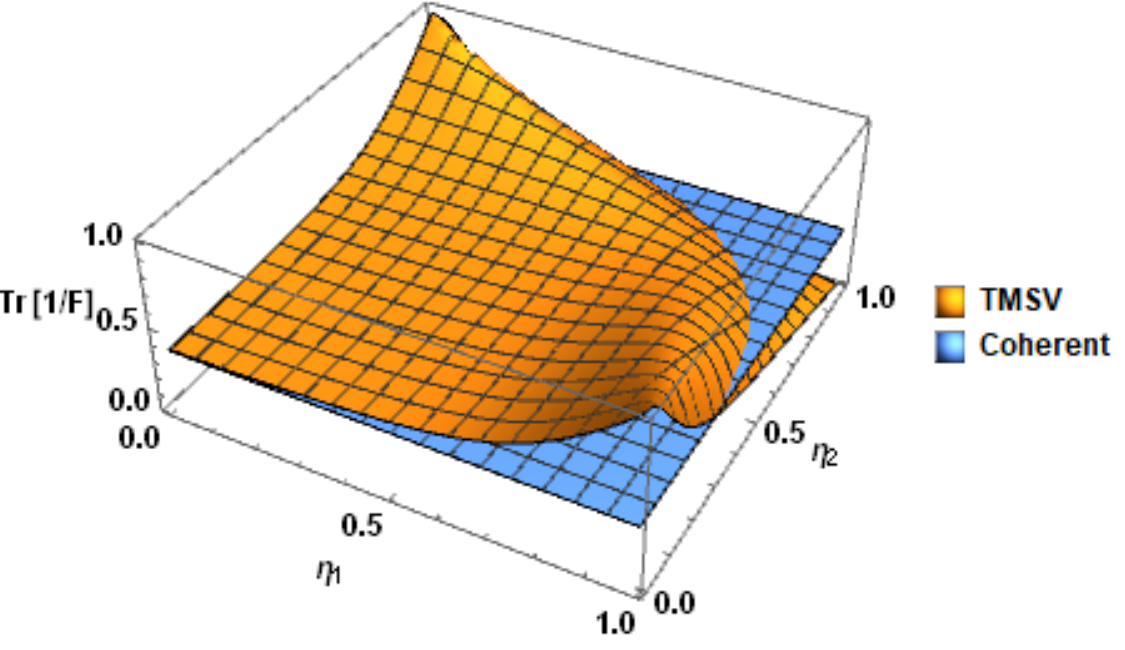}
    \caption{Total variance versus loss parameters $\eta_1$ and $\eta_2$ using PNRDs that can distinguish up to $5$ photons for a squeezed state with $r=1$. The sensitivity for TMSV is significantly better than that of coherent states when loss is small, now levels of loss with transmission amplitudes $\eta_i\approx 0.8$ negate the quantum advantage, and TMSV is worse than classical states when the loss is highly asymmetric.}
    \label{fig:QFI TMSV Xanadu vs coh PNRD inverse}
\end{figure}

\begin{figure}
    \centering
    \includegraphics[width=\columnwidth]{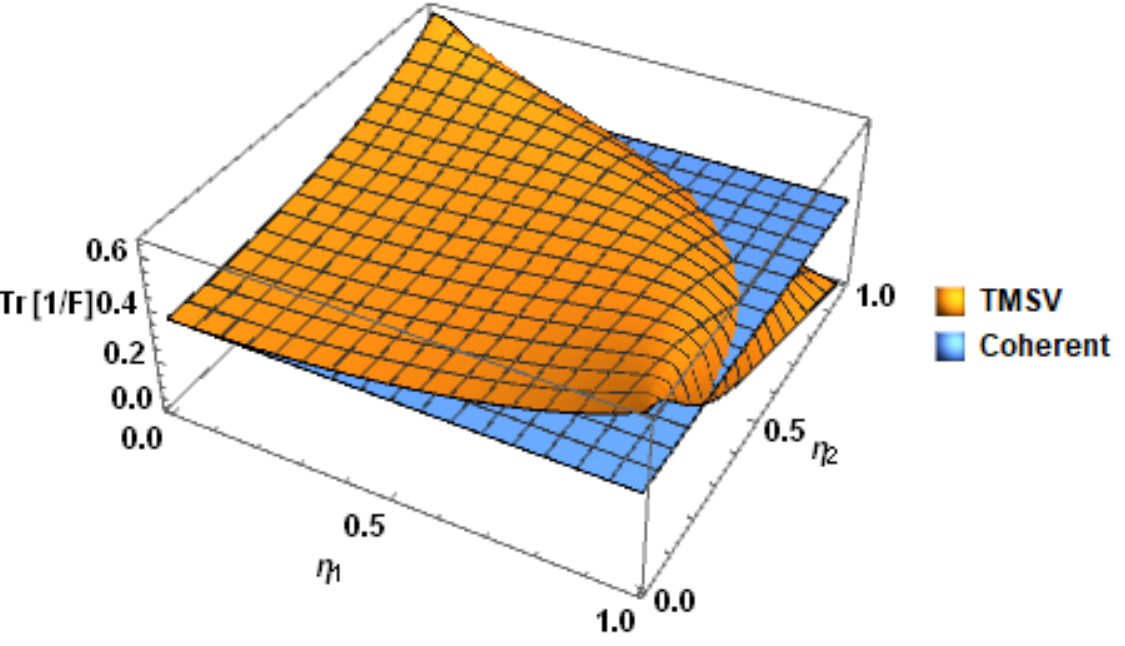}
    \caption{Total variance versus loss parameters $\eta_1$ and $\eta_2$ using PNRDs that can distinguish up to $10$ photons for a squeezed state with $r=1$. The sensitivity for TMSV is significantly better than that of coherent states when loss is small, now levels of loss with transmission amplitudes $\eta_i\approx 0.8$ negate the quantum advantage, and TMSV is worse than classical states when the loss is highly asymmetric.}
    \label{fig:QFI TMSV Xanadu vs coh P10NRD inverse}
\end{figure}

\begin{figure}
    \centering
    \includegraphics[width=\columnwidth]{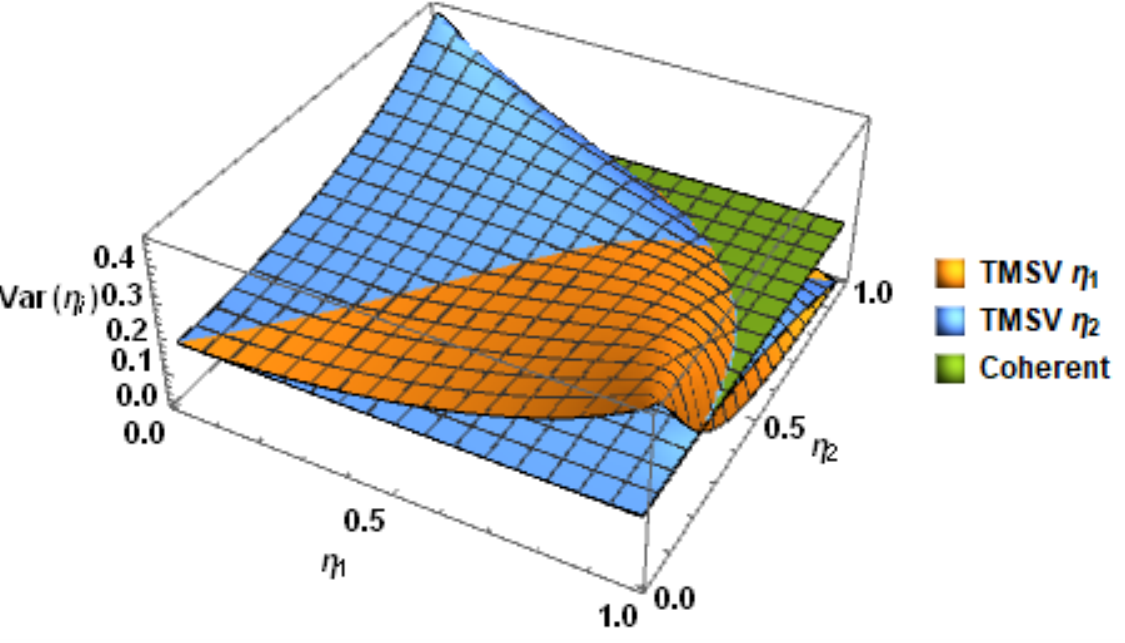}
    \caption{Specific variance of each loss parameter versus loss parameters $\eta_1$ and $\eta_2$ using PNRDs that can distinguish up to $10$ photons for a squeezed state with $r=1$. The sensitivity for TMSV is significantly better than that of coherent states when loss is small, now levels of loss with transmission amplitudes $\eta_i\approx 0.8$ negate the quantum advantage, and TMSV is worse than classical states when the loss is highly asymmetric.}
    \label{fig:uncertainty each param}
\end{figure}

The earlier figures displayed how TMSV outperforms coherent states in the large-transmission limit; now we inspect the large-loss limit.
By plotting the total variance $\mathrm{Tr}(1/\mathsf{\mathbf{F}})$ instead of the sensitivity $1/\mathrm{Tr}(1/\mathsf{\mathbf{F}})$, we can inspect how coherent states outperform TMSV in the large-loss limit. Figures 
\ref{fig:QFI TMSV Xanadu vs coh PNRD inverse} and \ref{fig:QFI TMSV Xanadu vs coh P10NRD inverse} demonstrate this total variance for PNRDs capable of resolving up to 5 and 10 photons, respectively, where it is clear that
TMSV and coherent states perform comparably when both modes have a large amount of loss, while coherent states outperform TMSV when one mode has a huge amount of loss and the other a small amount. Figure \ref{fig:uncertainty each param} unpacks the uncertainty attributed to the estimate of each parameter, where it is clear that having large transmission in mode 1 and large loss in mode 2 leads to a poor estimate of the transmission parameter of mode 2; high transmission in one mode is required to better estimate the transmission parameter in the other mode.


In the results section, we only featured plots for the first mode pair. We use this section to plot the same data for the other mode pairs.

\begin{figure}
    \centering
    \includegraphics[width=\columnwidth]{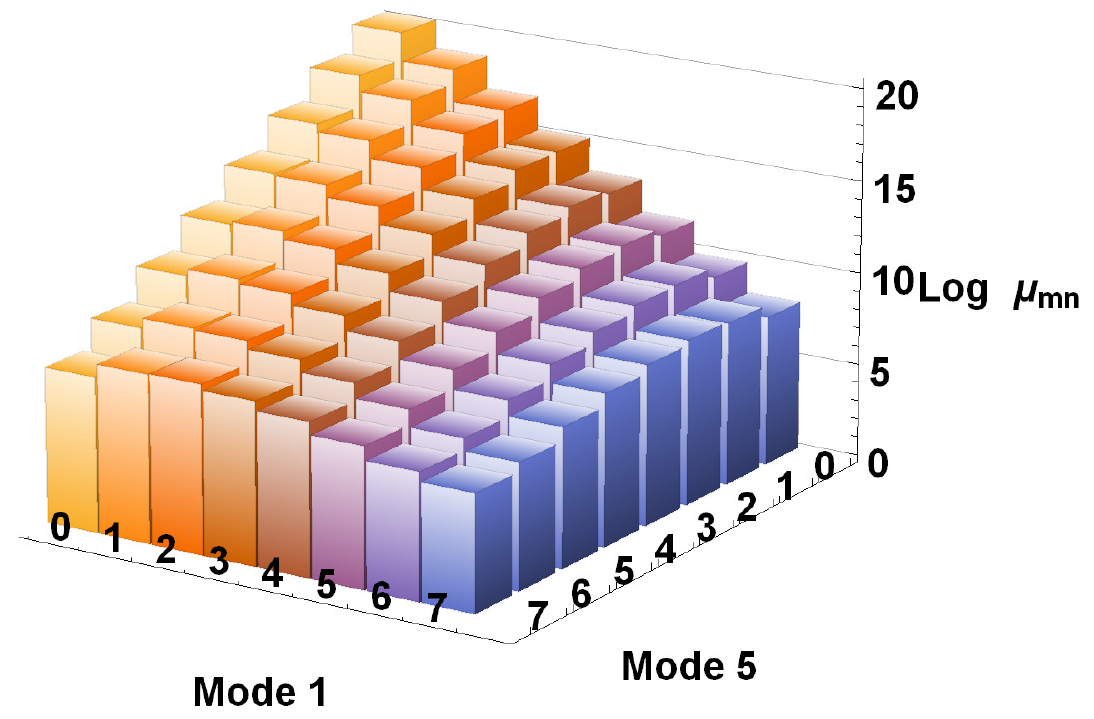}
    \caption{Semilog plot of the average measured photon-number distribution for TMSV states in modes 1 and 5 on Xanadu's X8 chip.}
    \label{fig:Log histograms 15}
\end{figure}

\begin{figure}
    \centering
    \includegraphics[width=\columnwidth]{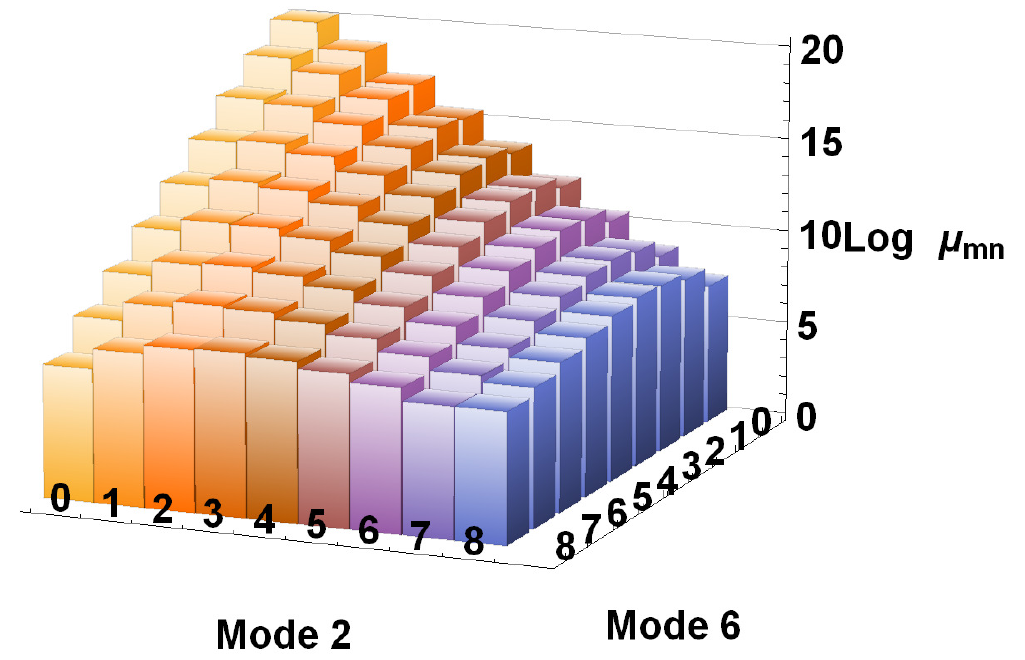}
    \caption{Semilog plot of the average measured photon-number distribution for TMSV states in modes 2 and 6 on Xanadu's X8 chip.}
    \label{fig:Log histograms 26}
\end{figure}

\begin{figure}
    \centering
    \includegraphics[width=\columnwidth]{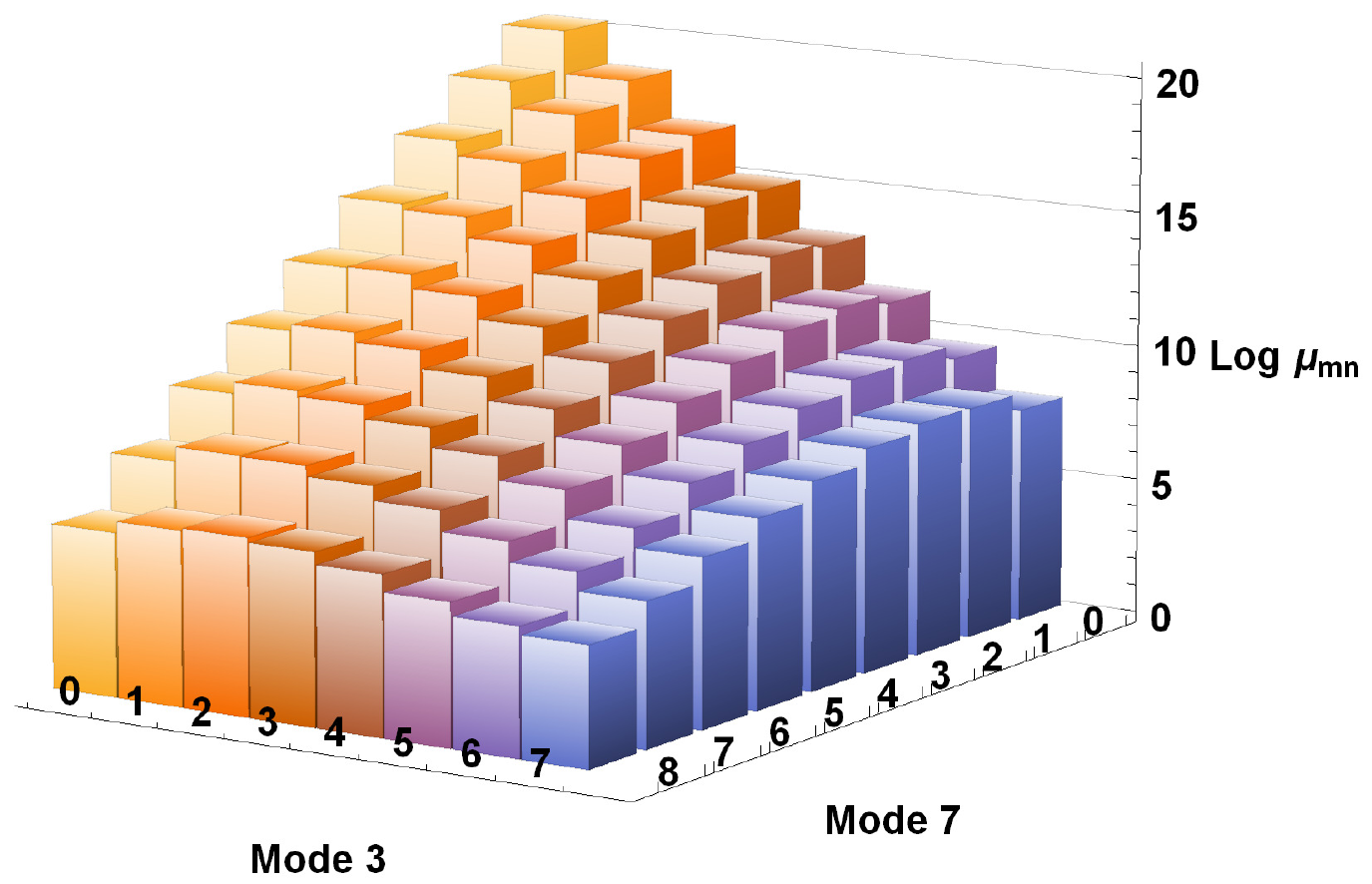}
    \caption{Semilog plot of the average measured photon-number distribution for TMSV states in modes 3 and 7 on Xanadu's X8 chip.}
    \label{fig:Log histograms 37}
\end{figure}

The observed photon-number distributions from the remaining three pairs of TMSV states in mode pairs 1 and 5, 2 and 6, and 3 and 7 are plotted in Figs. \ref{fig:Log histograms 15},  \ref{fig:Log histograms 26}, and  \ref{fig:Log histograms 37}, respectively.

\begin{figure}
    \centering
    \includegraphics[width=\columnwidth]{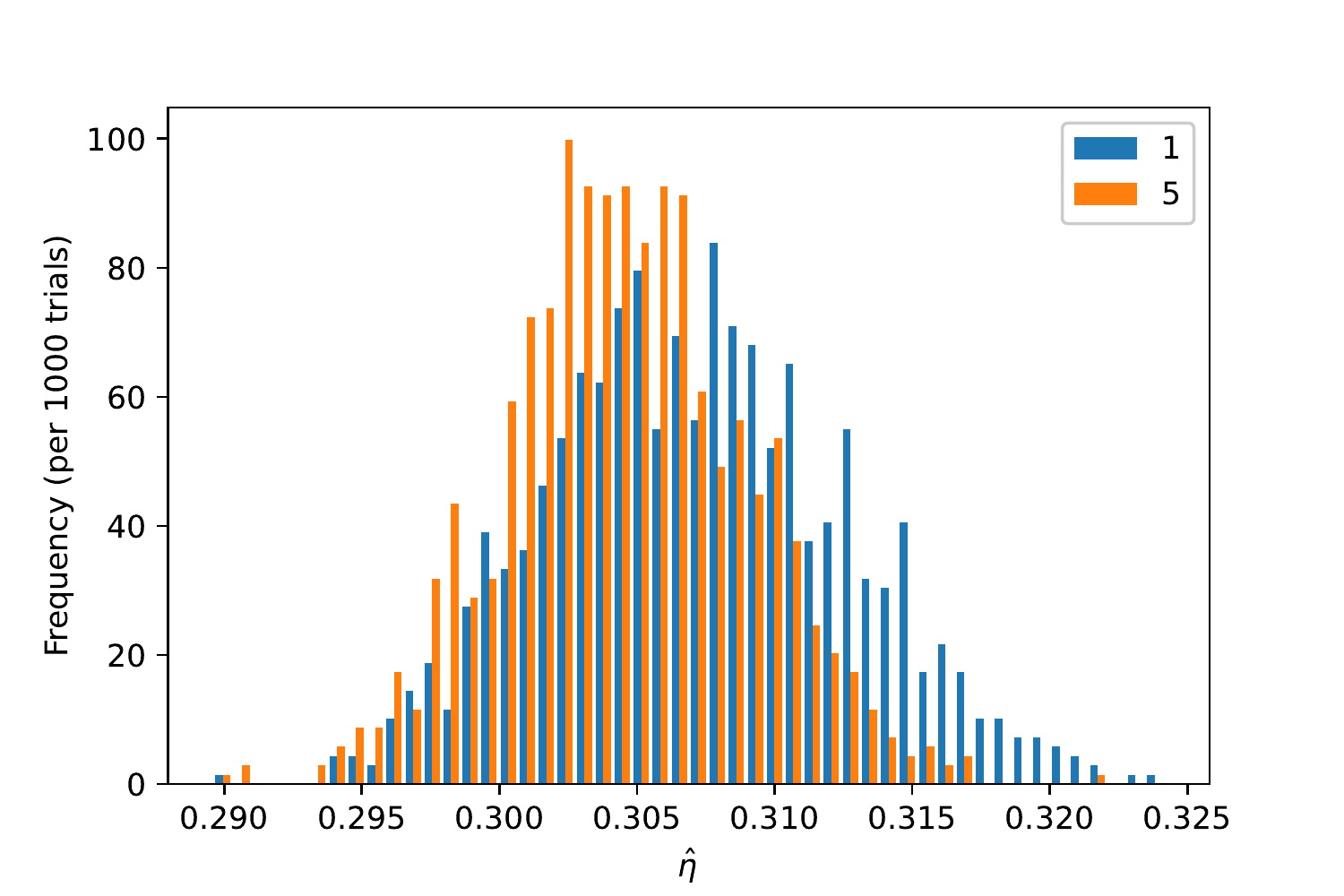}
    \caption{Estimated transmission parameters for each of the $10^3$ trials with TMSV in the pair of modes 1 and 5. The mean values of these data are 0.307(5) and 0.305(4), with the first digit of their standard deviations shown in parentheses. This corresponds to transmission probabilities of $\eta_1^2\approx 0.0942$ and $\eta_2^2\approx 0.0930$.}
    \label{fig:estimator hist eta 15 5par}
\end{figure}

\begin{figure}
    \centering
    \includegraphics[width=\columnwidth]{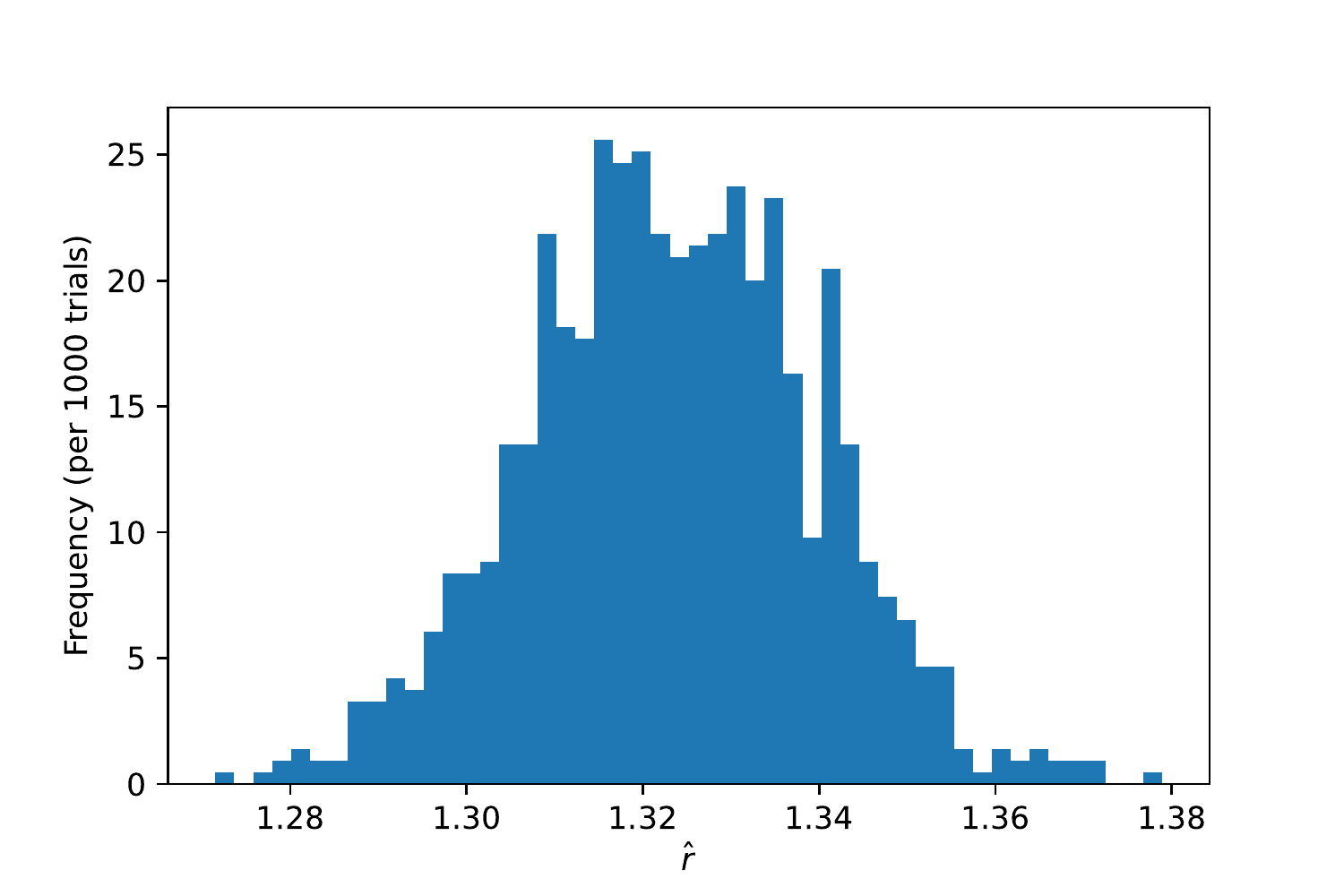}
    \caption{Estimated effective squeezing parameter for each of the $10^3$ trials with TMSV in the pair of modes 1 and 5. The mean value of these data is 1.32(2), with the first digit of its standard deviation shown in parentheses. This corresponds to an average photon number in the generated TMSV state of $2\sinh^2 r\approx 6.04$.}
    \label{fig:estimator hist r 15 5par}
\end{figure}

\begin{figure}
    \centering
    \includegraphics[width=\columnwidth]{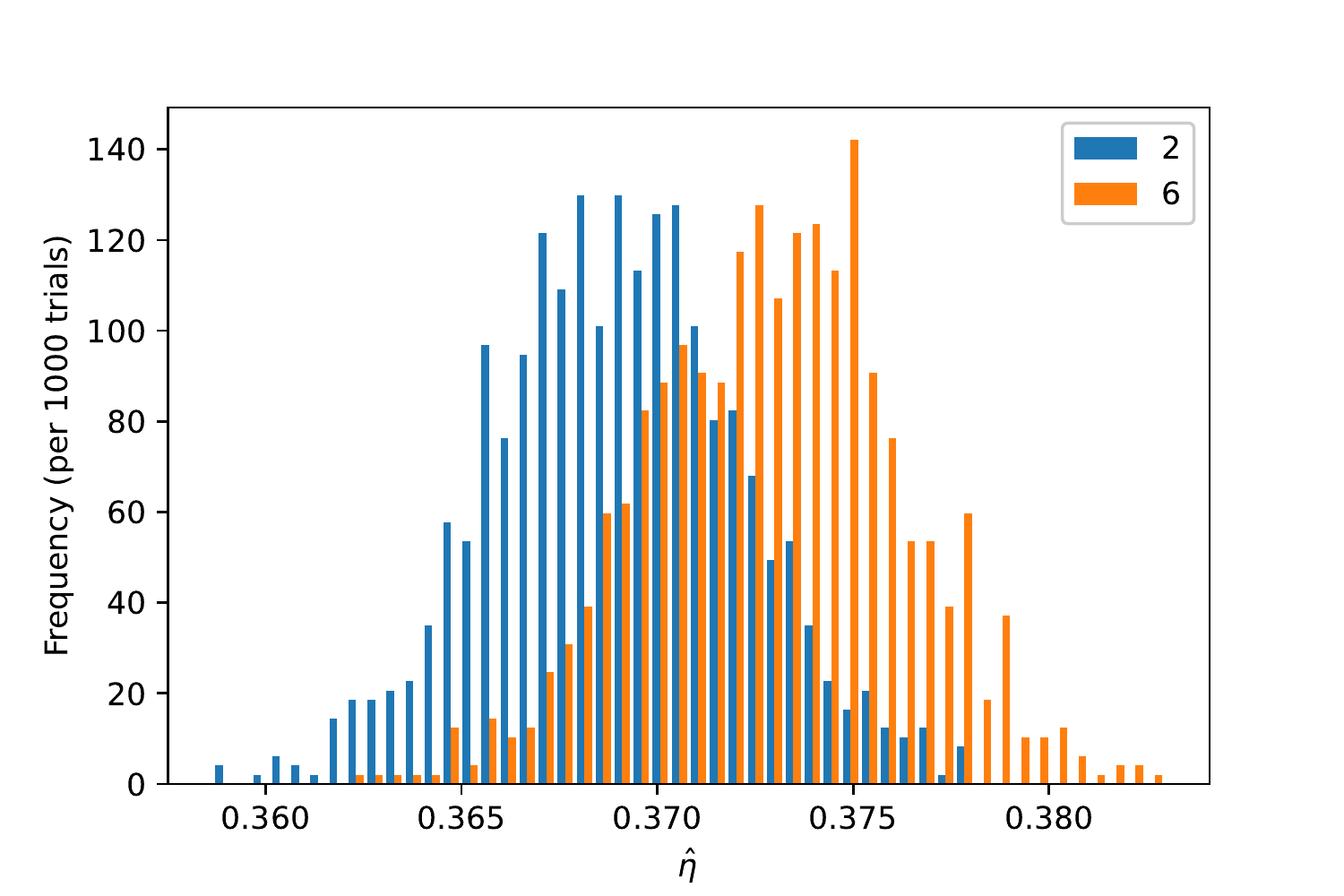}
    \caption{Estimated transmission parameters for each of the $10^3$ trials with TMSV in the pair of modes 2 and 6. The mean values of these data are 0.369(3) and 0.373(3), with the first digit of their standard deviations shown in parentheses. This corresponds to transmission probabilities of $\eta_1^2\approx 0.136$ and $\eta_2^2\approx 0.139$.}
    \label{fig:estimator hist eta 26 5par}
\end{figure}

\begin{figure}
    \centering
    \includegraphics[width=\columnwidth]{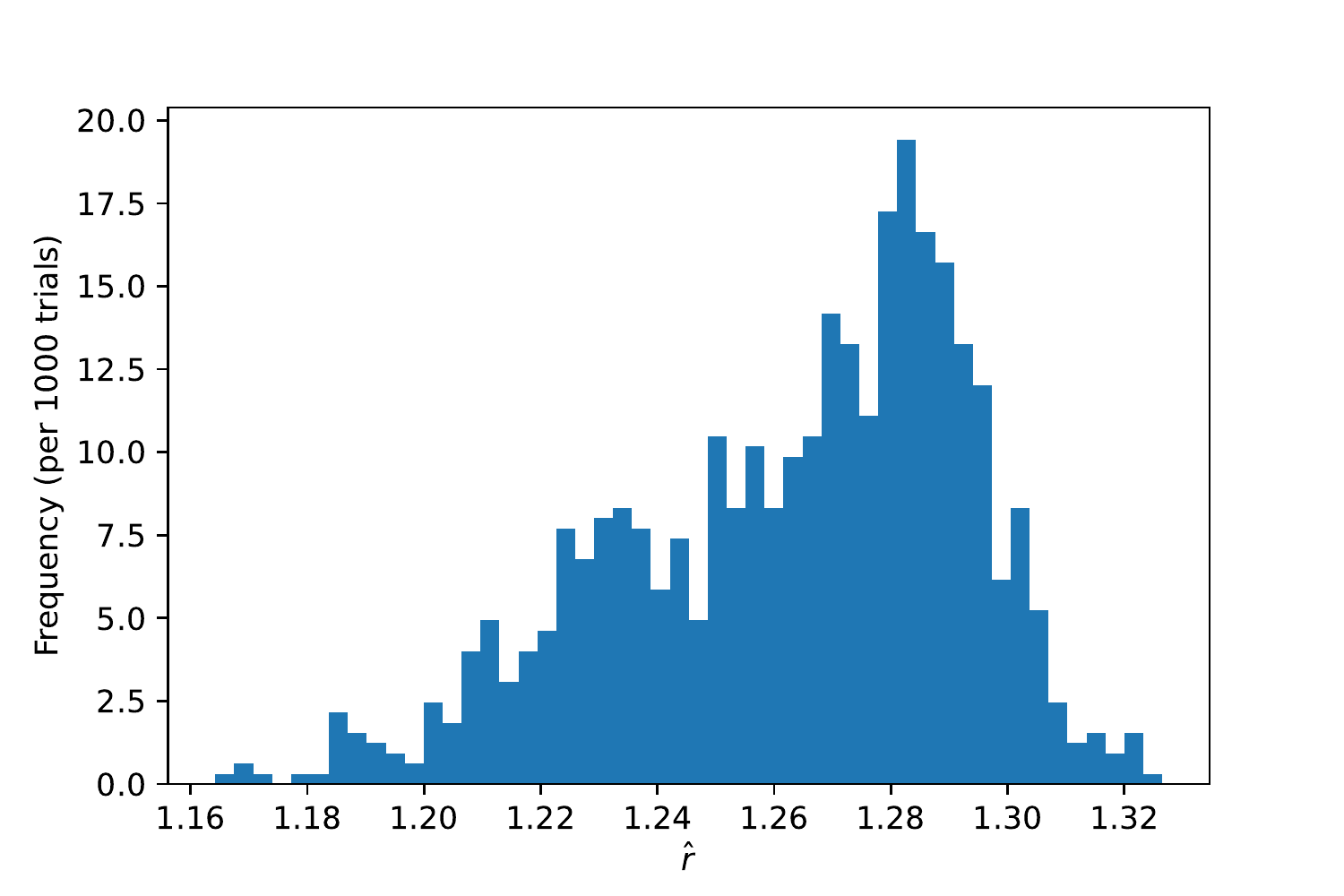}
    \caption{Estimated effective squeezing parameter for each of the $10^3$ trials with TMSV in the pair of modes 2 and 6. The mean value of these data is 1.26(3), with the first digit of its standard deviation shown in parentheses. This corresponds to an average photon number in the generated TMSV state of $2\sinh^2 r\approx 5.25$.}
    \label{fig:estimator hist r 26 5par}
\end{figure}

\begin{figure}
    \centering
    \includegraphics[width=\columnwidth]{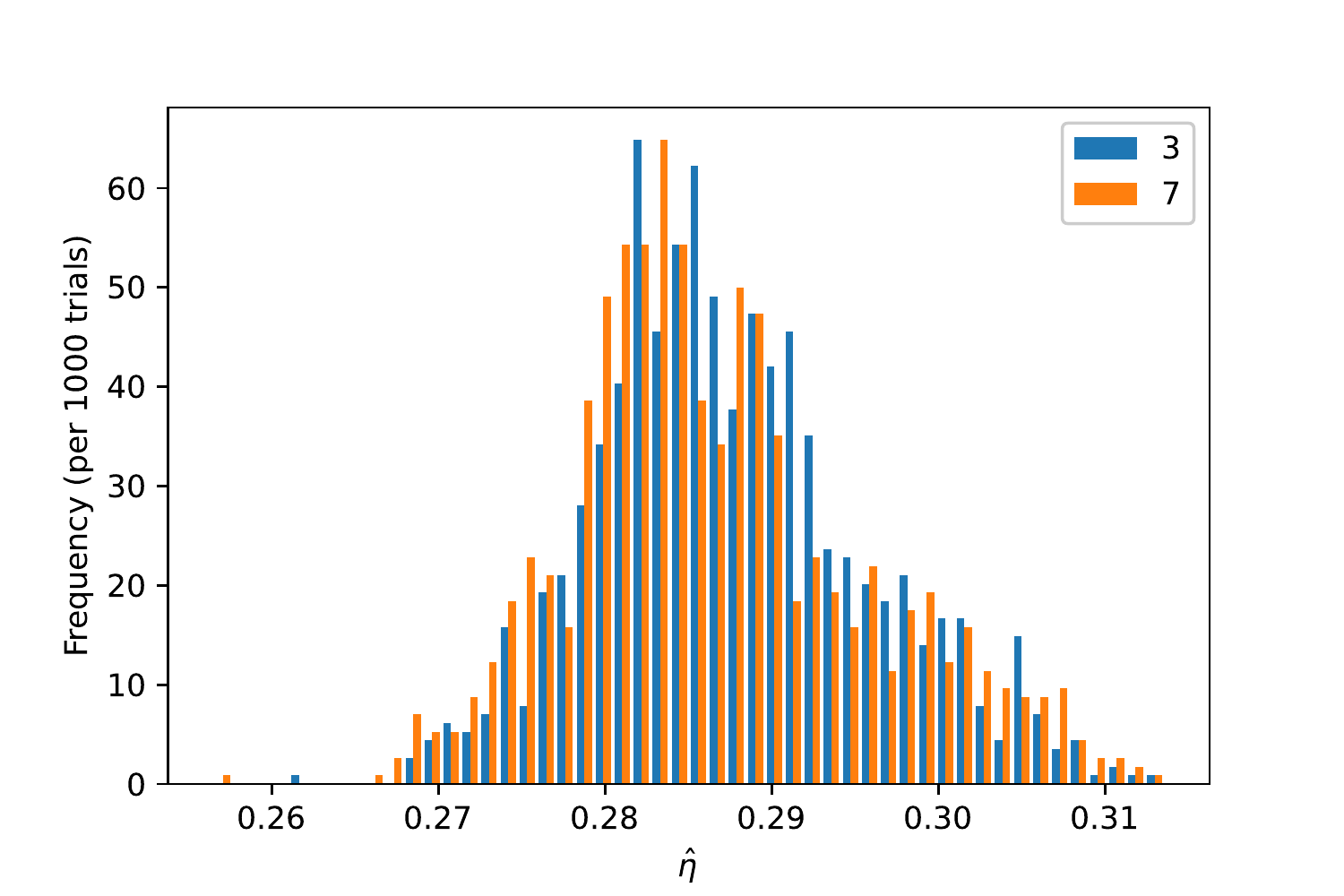}
    \caption{Estimated transmission parameters for each of the $10^3$ trials with TMSV in the pair of modes 3 and 7. The mean values of these data are 0.288(8) and 0.287(9), with the first digit of their standard deviations shown in parentheses. This corresponds to transmission probabilities of $\eta_1^2\approx 0.0829$ and $\eta_2^2\approx 0.0824$.}
    \label{fig:estimator hist eta 37 5par}
\end{figure}

\begin{figure}
    \centering
    \includegraphics[width=\columnwidth]{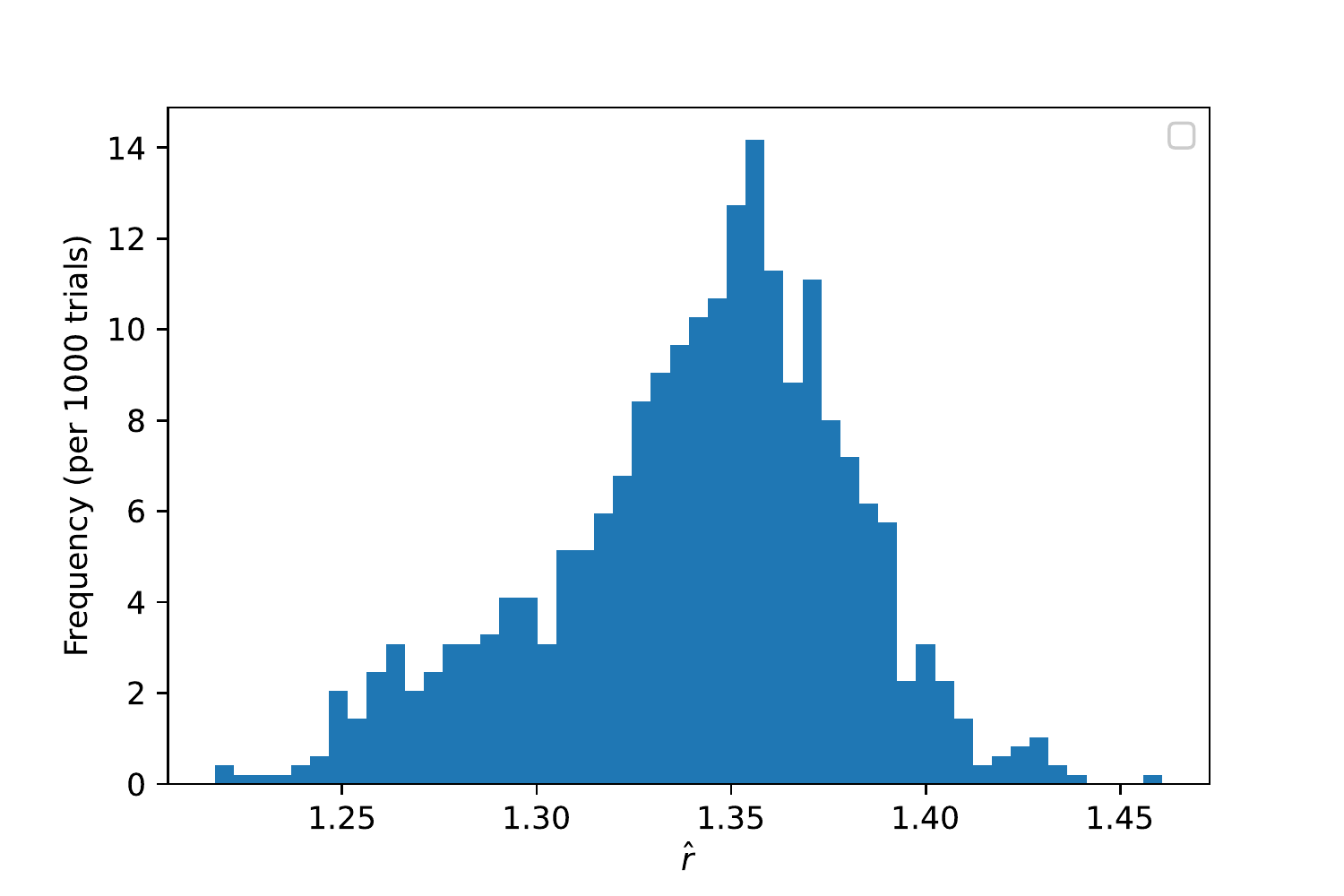}
    \caption{Estimated effective squeezing parameter for each of the $10^3$ trials with TMSV in the pair of modes 3 and 7. The mean value of these data is 1.34(4), with the first digit of its standard deviation shown in parentheses. This corresponds to an average photon number in the generated TMSV state of $2\sinh^2 r\approx 6.33$.}
    \label{fig:estimator hist r 37 5par}
\end{figure}

The histograms of the MLE results for the TMSVs in mode pairs 1 and 5, 2 and 6, and 3 and 7 in Figs. \ref{fig:estimator hist eta 15 5par} and \ref{fig:estimator hist r 15 5par}, \ref{fig:estimator hist eta 26 5par} and \ref{fig:estimator hist r 26 5par}, and \ref{fig:estimator hist eta 37 5par} and \ref{fig:estimator hist r 37 5par}, respectively.



\begin{figure}
    \centering
    \includegraphics[width=\columnwidth]{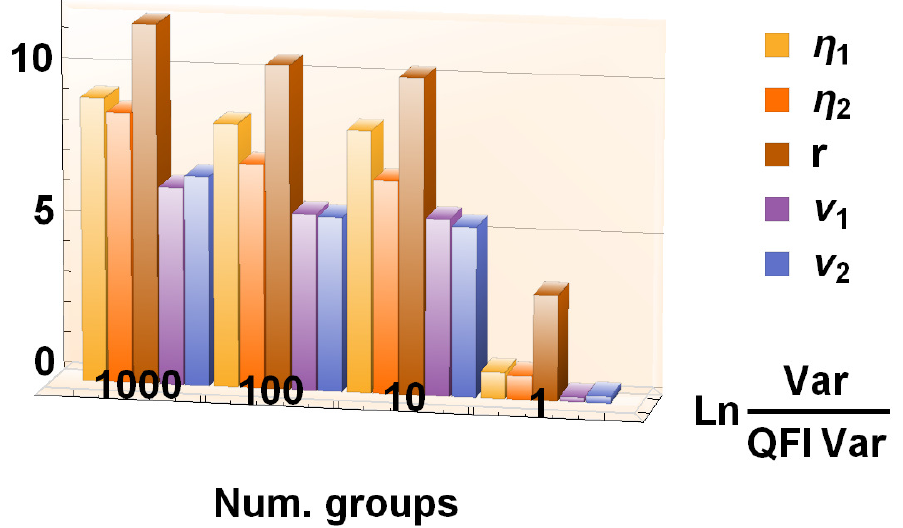}
    \caption{Variances of the MLE estimates from the data aggregated into different numbers of groups, relative to the theoretical limit of what can be obtained using TMSV, for modes 1 and 5. The variances of of the parameters hardly change, demonstrating that more shots within a trial is no longer superior to having more total trials with fewer shots for this data set. The QFI value for the variances of the dark count parameters $\hat{\nu}_i$ is artificially set to $10^{-9}$ for scaling purposes.}
    \label{fig:var change size 15 5par}
\end{figure}

\begin{figure}
    \centering
    \includegraphics[width=\columnwidth]{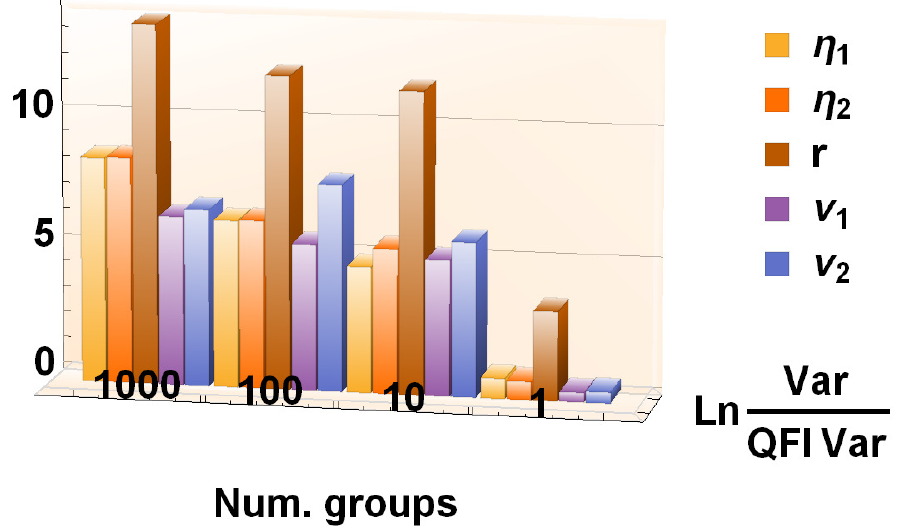}
    \caption{Variances of the MLE estimates from the data aggregated into different numbers of groups, relative to the theoretical limit of what can be obtained using TMSV, for modes 2 and 6. The variances of $\hat{\eta}_1$ and $\hat{\eta}_2$ diminish with increasing group size (decreasing number of groups), while the variances of the other estimated parameters decrease nonmonotonically, demonstrating that more shots within a trial is superior to having more total trials with fewer shots. The QFI value for the variances of the dark count parameters $\hat{\nu}_i$ is artificially set to $10^{-9}$ for scaling purposes.}
    \label{fig:var change size 26 5par}
\end{figure}

\begin{figure}
    \centering
    \includegraphics[width=\columnwidth]{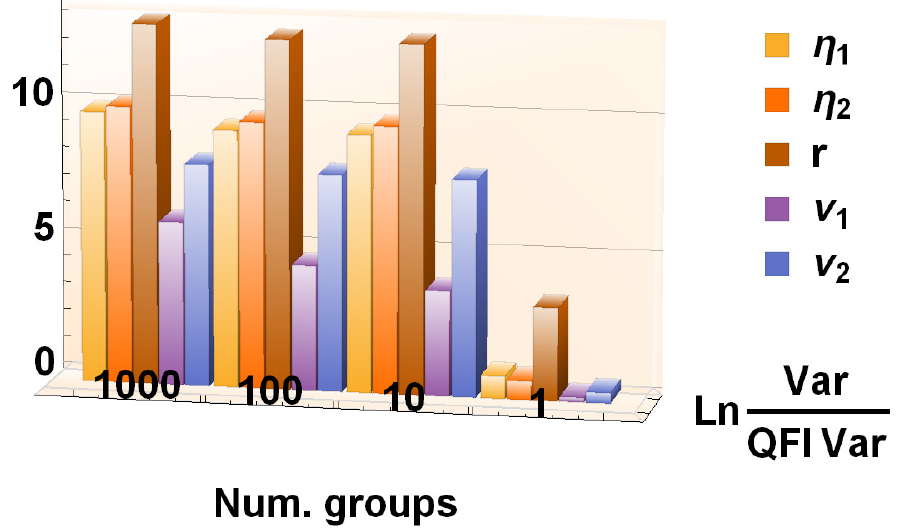}
    \caption{Variances of the MLE estimates from the data aggregated into different numbers of groups, relative to the theoretical limit of what can be obtained using TMSV, for modes 3 and 7. The variances of of the parameters hardly change, demonstrating that more shots within a trial is no longer superior to having more total trials with fewer shots for this data set. The QFI value for the variances of the dark count parameters $\hat{\nu}_i$ is artificially set to $10^{-9}$ for scaling purposes.}
    \label{fig:var change size 37 5par}
\end{figure}

Then, we plot the variances of the estimates after aggregating the data into different group sizes, including our estimates of the dark count rates. Modes 1 and 5, 2 and 6, and 3 and 7 are plotted in Figs. \ref{fig:var change size 15 5par}, \ref{fig:var change size 26 5par}, and \ref{fig:var change size 37 5par} respectively.

\begin{figure}
    \centering
    \includegraphics[width=\columnwidth]{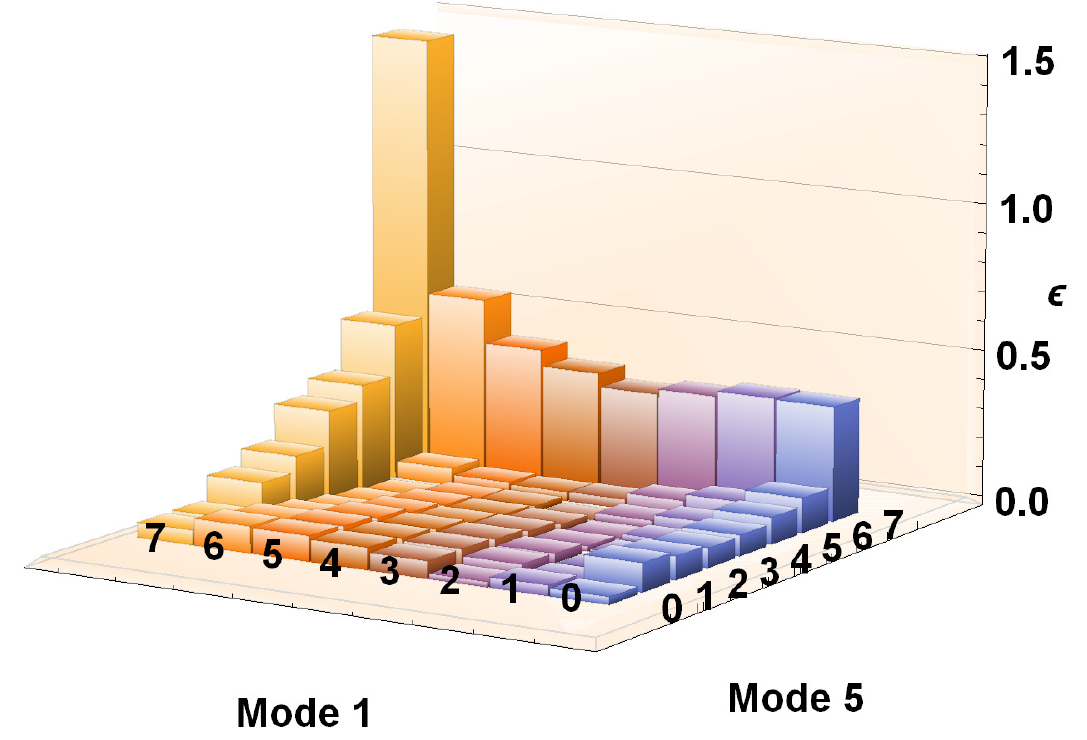}
    \caption{Absolute relative errors between the observed frequencies and expected probabilities for each photon-number-detection event in modes 0 and 4. These relative errors are sufficiently small to justify using Eq. \eqref{eq:FI from observed likelihood} to estimate the covariances of the underlying parameters. The average value is 0.129186. Note that the order of the photon numbers is opposite relative to Fig. \ref{fig:Log histograms 15} to make the data more visible.}
    \label{fig:rel err15 5par}
\end{figure}

\begin{figure}
    \centering
    \includegraphics[width=\columnwidth]{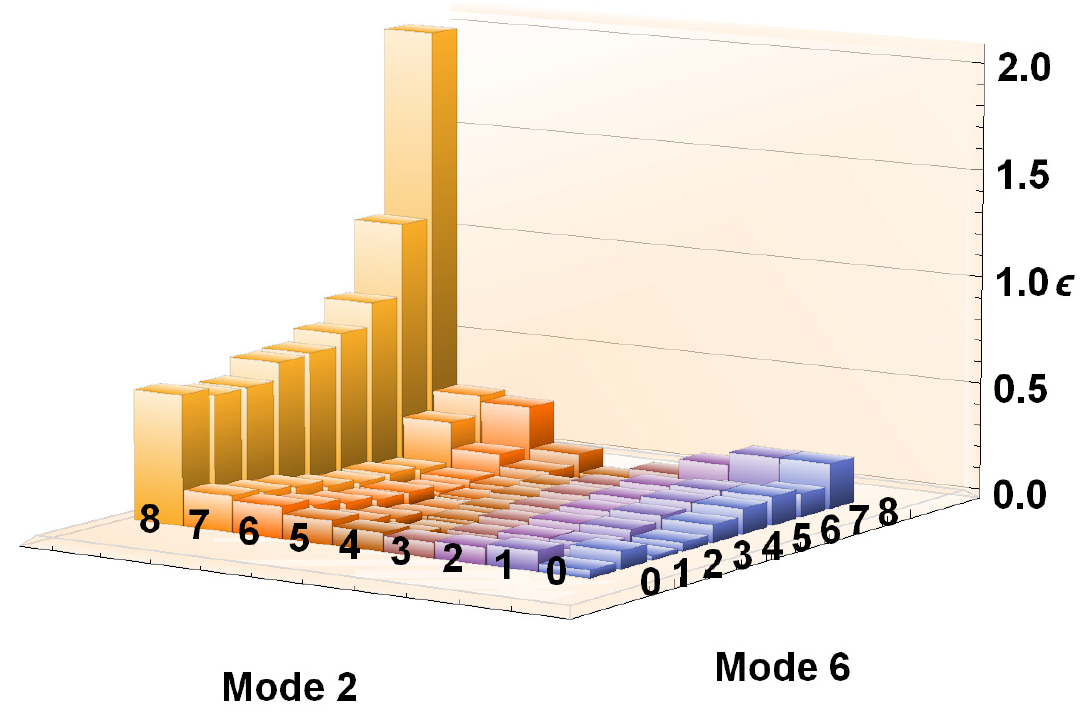}
    \caption{Absolute relative errors between the observed frequencies and expected probabilities for each photon-number-detection event in modes 0 and 4. These relative errors are sufficiently small to justify using Eq. \eqref{eq:FI from observed likelihood} to estimate the covariances of the underlying parameters. The average value is 0.170125. Note that the order of the photon numbers is opposite relative to Fig. \ref{fig:Log histograms 26} to make the data more visible.}
    \label{fig:rel err26 5par}
\end{figure}

\begin{figure}
    \centering
    \includegraphics[width=\columnwidth]{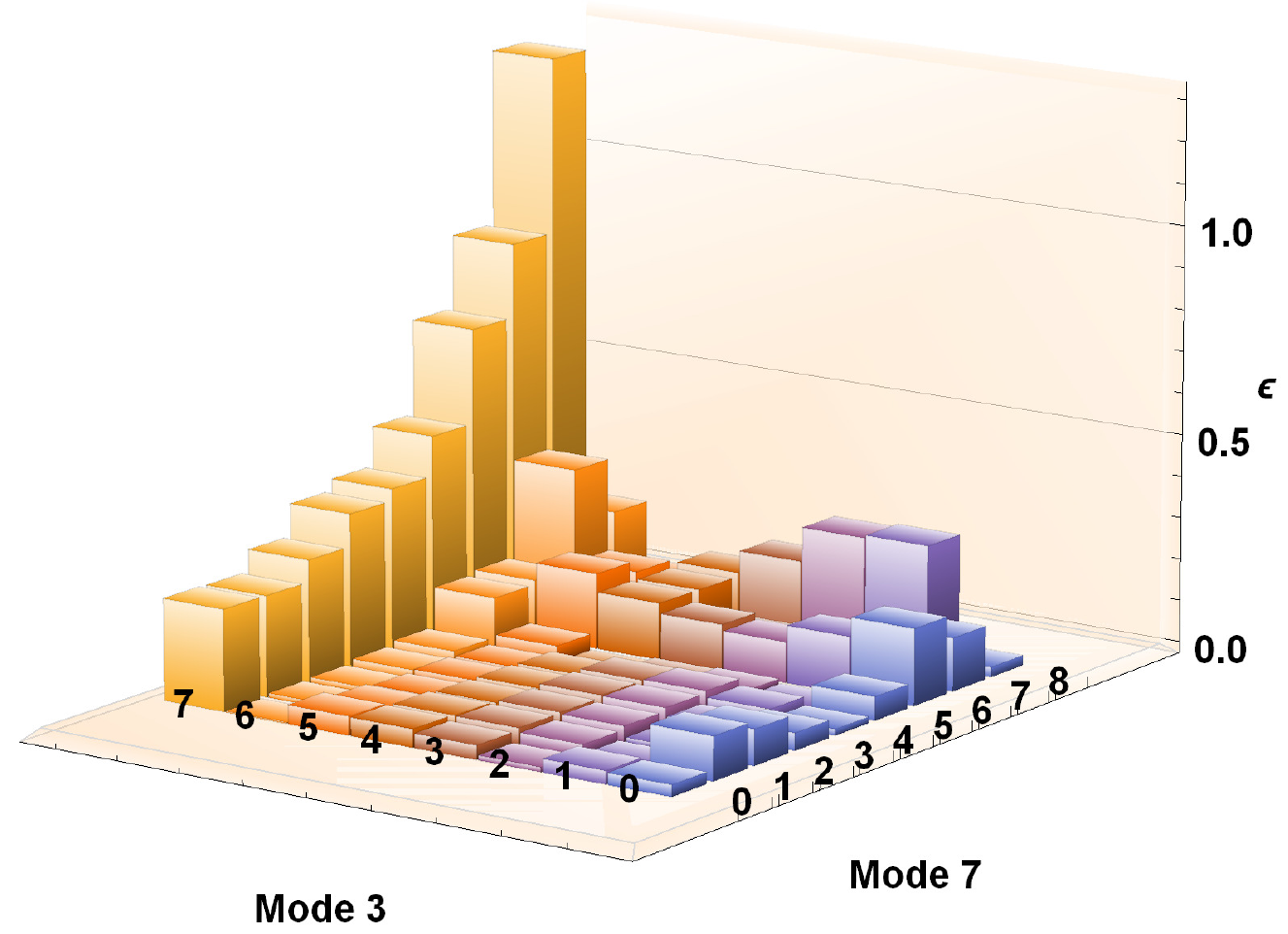}
    \caption{Absolute relative errors between the observed frequencies and expected probabilities for each photon-number-detection event in modes 0 and 4. These relative errors are sufficiently small to justify using Eq. \eqref{eq:FI from observed likelihood} to estimate the covariances of the underlying parameters. The average value is 0.133507. Note that the order of the photon numbers is opposite relative to Fig. \ref{fig:Log histograms 37} to make the data more visible.}
    \label{fig:rel err37 5par}
\end{figure}

Lastly, we plot the relative errors for the various mode pairs after aggregating all of the data in Figs. \ref{fig:rel err15 5par}, \ref{fig:rel err26 5par}, and \ref{fig:rel err37 5par}.

	\clearpage
\section{Gaussian calculations}
\label{app:Gaussian calcs}

One can compute the QFIM directly using techniques from Gaussian states (e.g.,  Ref. \cite{Brask2021arxiv}). Gaussian states such as coherent states and TMSV are uniquely described by their first and second moments. For TMSV, the first moments vanish and the covariance matrix takes the form (we take $\hbar=1$)
\eq{
    \pmb{\sigma}=\frac{1}{2}\begin{pmatrix}
    \mathds{1}\cosh 2r&-\mathbf{S}_\varphi\sinh 2r\\
    -\mathbf{S}_\varphi\sinh 2r &\mathds{1}\cosh 2r
    \end{pmatrix},
} where $\mathbf{S}_\varphi=\begin{pmatrix}\cos\varphi&\sin\varphi\\\sin\varphi&-\cos\varphi\end{pmatrix}$ is not quite a rotation matrix and we are using squeezing operators $\exp(\xi^* \ha\hb-\xi \had\hbd)$ with $\xi=r \eu^{\iu\varphi}$. The basis is $(x_1,p_1,x_2,p_2)$, where $x=(\ha+\had)/\sqrt{2}$, $p=(\ha-\had)/\sqrt{2}\iu$, and $[x,p]=\iu$. Loss can be enacted by a beam splitter with initially unpopulated modes and then the neglect of those additional modes, so we have to expand our covariance matrix to
\eq{
    \pmb{\sigma}=\frac{1}{2}\begin{pmatrix}
    \mathds{1}\cosh 2r&\mathbf{0}&-\mathbf{S}_\varphi\sinh 2r&\mathbf{0}\\
    \mathbf{0}&\mathds{1}&\mathbf{0}&\mathbf{0}\\
    -\mathbf{S}_\varphi\sinh 2r&\mathbf{0} &\mathds{1}\cosh 2r&\mathbf{0}\\
    \mathbf{0}&\mathbf{0}&\mathbf{0}&\mathds{1}
    \end{pmatrix}.
} For the beam splitters, which for two modes are represented by $\mathbf{F}=\begin{pmatrix}\eta\mathds{1}&\sqrt{1-\eta^2}\mathds{1}\\
-\sqrt{1-\eta^2}\mathds{1}&\eta\mathds{1}\end{pmatrix}$, we act with a matrix $\mathbf{F}(\eta_i)$ on both sides, for each $\eta_i$, to find
\begin{widetext}
\begin{equation}
    \pmb{\sigma}=\frac{1}{2}\left(
\begin{array}{cccc}
 (c-1) \eta_1^2+1 & -(c-1) \eta_1 \sqrt{1-\eta_1^2} & -\eta_1 \eta_2 s \mathbf{S}_\varphi & \eta_1 \sqrt{1-\eta_2^2} s \mathbf{S}_\varphi \\
 -(c-1) \eta_1 \sqrt{1-\eta_1^2} & -c \eta_1^2+\eta_1^2+c & \sqrt{1-\eta_1^2} \eta_2 s \mathbf{S}_\varphi & -\sqrt{1-\eta_1^2} \sqrt{1-\eta_2^2} s \mathbf{S}_\varphi \\
 -\eta_1 \eta_2 s \mathbf{S}_\varphi & \sqrt{1-\eta_1^2} \eta_2 s \mathbf{S}_\varphi & (c-1) \eta_2^2+1 & -(c-1) \eta_2 \sqrt{1-\eta_2^2} \\
 \eta_1 \sqrt{1-\eta_2^2} s \mathbf{S}_\varphi & -\sqrt{1-\eta_1^2} \sqrt{1-\eta_2^2} s \mathbf{S}_\varphi & -(c-1) \eta_2 \sqrt{1-\eta_2^2} & -c \eta_2^2+\eta_2^2+c \\
\end{array}
\right).
\end{equation}
\end{widetext} 
Then, performing the partial trace to ignore the vacuum modes is the same as deleting those unnecessary rows and columns from this standard form, so our transformed covariance matrix is
\begin{equation}
    \pmb{\sigma}=\frac{1}{2}\begin{pmatrix}
 \left[(\cosh2r-1) \eta_1^2+1\right]\mathds{1} &  -\eta_1 \eta_2 \sinh2r \mathbf{S}_\varphi  \\
 -\eta_1 \eta_2 \sinh2r \mathbf{S}_\varphi  & \left[(\cosh2r-1) \eta_2^2+1\right]\mathds{1}  \end{pmatrix}.
\end{equation}
This is exactly the same covariance matrix as for our state given by Eq. \eqref{eq:rho after TMSV loss}, so either method could have been used to arrive at this result.

We can now calculate the QFIM directly from this covariance matrix using the techniques from Ref. \cite{Safranek2018}.
The expression is slightly cumbersome, but its inverse in the $(\eta_1,\eta_2,r)$-basis is less so:
\begin{widetext}
\eq{
    \mathsf{\mathbf{F}}^{-1}=
    \left(
\begin{array}{ccc}
 \frac{\left(\eta_1^2-1\right) \left(\eta_2^2-\text{csch}^2(r)-1\right)}{4 \eta_2^2} & \frac{\left(\eta_1^2-1\right) \left(\eta_2^2-1\right) \coth ^2(r)}{4 \eta_1 \eta_2} & -\frac{\left(\eta_1^2-1\right) \left(\eta_2^2-1\right) \coth (r)}{4 \eta_1 \eta_2^2} \\
 \frac{\left(\eta_1^2-1\right) \left(\eta_2^2-1\right) \coth ^2(r)}{4 \eta_1 \eta_2} & \frac{\left(\eta_2^2-1\right) \left(\eta_1^2-\text{csch}^2(r)-1\right)}{4 \eta_1^2} & -\frac{\left(\eta_1^2-1\right) \left(\eta_2^2-1\right) \coth (r)}{4 \eta_1^2 \eta_2} \\
 -\frac{\left(\eta_1^2-1\right) \left(\eta_2^2-1\right) \coth (r)}{4 \eta_1 \eta_2^2} & -\frac{\left(\eta_1^2-1\right) \left(\eta_2^2-1\right) \coth (r)}{4 \eta_1^2 \eta_2} & \frac{1}{4} \left(\frac{\frac{1}{\eta_2^2}-1}{\eta_1^2}-\frac{1}{\eta_2^2}+2\right) \\
\end{array}
\right).
}
\end{widetext}
The diagonal components of this inverse are what we quote in Eq. \eqref{eq:variance bounds TMSV nuisance}.

	\clearpage
	\section{Derivation of estimated Fisher information from observed likelihood Eq. \eqref{eq:FI from observed likelihood}}
	\label{app:FI from likelihood}
	We start with the definition of the Fisher information
	\eq{
    \mathsf{F}_{jk}&=\expct{\frac{\partial \ln \mathcal{L}}{\partial \eta_j}\frac{\partial \ln \mathcal{L}}{\partial \eta_k}},
    } where the expectation value is taken over the probability distribution for the possible observed events. The underlying probabilities are taken Each log-likelihood is given by a sum as in Eq. The underlying probabilities $p(m,n|\hat{\pmb{\theta}})$ are taken to be fixed while the observed counts $\mu_{mn}$ follow a multinomial distribution (``how many ways are there of arranging $N$ outcomes such that the event measuring $m$ and $n$ occurs on $\mu_{mn}$ occasions?'')
    \begin{widetext}
    \eq{
        \expct{\mu_{mn}\mu_{m^\prime n^\prime}}&=\sum_{x_{00}+x_{01}+\cdots+x_{kl}=N}\binom{N}{x_{00},x_{01},\cdots,x_{kl}}p(0,0|\hat{\pmb{\theta}})^{x_{00}} p(0,1|\hat{\pmb{\theta}})^{x_{01}}\cdots p(k,l|\hat{\pmb{\theta}})^{x_{kl}} x_{mn}x_{m^\prime n^\prime}\\
        &=\delta_{mm^\prime}\delta_{nn^\prime} \expct{\mu_{mn}}+p(m,n|\hat{\pmb{\theta}})p(m^\prime,n^\prime|\hat{\pmb{\theta}})(N^2-N).
    }
    \end{widetext}
    The second term vanishes in the Fisher information, because the expectation value of the score vanishes (the derivative of a constant total probability vanishes)
    \eq{
        \sum_{mn}p(m,n|\hat{\pmb{\theta}})\frac{\partial \ln p(m,n|\hat{\pmb{\theta}})}{\partial \eta_j}=0,
    } so we are left with
    \eq{
    \mathsf{F}_{jk}
    &=\sum_{mn}\expct{\mu_{mn}}\frac{\partial \ln p(m,n|\hat{\pmb{\theta}})}{\partial \eta_j}\frac{\partial \ln p(m,n|\hat{\pmb{\theta}})}{\partial \eta_k}.
} Assuming the observed data $\mu_{mn}$ to well represent the average data $\expct{\mu_{mn}}=N p(m,n|\hat{\pmb{\theta}})$, which holds true in the limit of large numbers of experiments, then yields Eq. \eqref{eq:FI from observed likelihood}.
	
	An alternative derivation considers the expectation value to be averaging over the possible photon-number-detection events labeled by $(m,n)$. Then the Fisher information for a single measurement has the components
	\eq{
	    \mathsf{F}_{jk}=\sum_{mn}p_{mn}\frac{\partial \ln p_{mn}}{\partial \eta_j}\frac{\partial \ln p_{mn}}{\partial \eta_k}.
	} Recalling that the underlying probabilities depend on the true parameters $\pmb{\theta}\approx \hat{\pmb{\theta}}$, scaling the Fisher information to include all $\mu$ measurements, and assuming that the measured distribution $\{\mu_{mn}\}$ well approximates the underlying probability distribution again yields Eq. \eqref{eq:FI from observed likelihood}
    \eq{
	    \mathsf{F}_{jk}=\sum_{mn}\mu_{mn}\frac{\partial \ln p(m,n|\hat{\pmb{\theta}})}{\partial \eta_j}\frac{\partial \ln p(m,n|\hat{\pmb{\theta}})}{\partial \eta_k}.
	} This is also equivalent to each measurement event $(m,n)$ contributing a Fisher information $\frac{\partial \ln p(m,n|\hat{\pmb{\theta}})}{\partial \eta_j}\frac{\partial \ln p(m,n|\hat{\pmb{\theta}})}{\partial \eta_k}$, since the Fisher information is additive for independent measurements.
\end{document}